\documentclass[12pt]{article}
\usepackage{iftex}
\ifPDFTeX\usepackage[T1]{fontenc}\fi
\usepackage{kpfonts}
\usepackage{amsmath,amsthm,mathtools}
\usepackage{enumitem}
\usepackage[margin=1.15in]{geometry}
\usepackage{setspace,microtype,xurl}
\usepackage{graphicx,booktabs,tabularx,array}
\usepackage[round,authoryear]{natbib}
\usepackage[section]{placeins}
\usepackage[colorlinks=true,linkcolor=blue,citecolor=blue,urlcolor=blue]{hyperref}
\allowdisplaybreaks
\newtheorem{theorem}{Theorem}
\newtheorem{proposition}{Proposition}
\newtheorem{lemma}{Lemma}
\newtheorem{corollary}{Corollary}
\newtheorem{assumption}{Assumption}
\theoremstyle{remark}

\theoremstyle{plain}
\newcommand{\E}{\mathbb E}
\newcommand{\Cov}{\operatorname{Cov}}

\newcommand{\dd}{\mathrm d}
\newcommand{\CS}{\mathrm{CS}}
\newcommand{\Ads}{\mathcal A}

\title{Platform Recommendation and Competition in Search Markets}
\author{Sanxi Li\thanks{School of Economics, Renmin University of China. sanxi@ruc.edu.cn} \and Jun Yu\thanks{School of Economics, Shanghai University of Finance and Economics. yu.jun@mail.shufe.edu.cn} \and Mingsheng Zhang\thanks{School of Economics, Renmin University of China. mingsheng\_zhang@yeah.net}}
\date{November 2026}
\hypersetup{pdftitle={Platform Recommendation and Competition in Search Markets},pdfauthor={Sanxi Li, Jun Yu, Mingsheng Zhang}}
\begin{document}
\maketitle
\begin{abstract}
Recommendation rules affect both sellers' incentives and the alternatives consumers expect to find. We develop a framework for evaluating these rules with heterogeneous sellers, endogenous pricing and advertising, and costly consumer search. For a broad class of separable recommendation scores, we establish conditions under which a unique market equilibrium exists. Three applications illustrate the use of the framework. A stronger reward for low prices does not necessarily lower equilibrium prices and benefit consumers when products are sufficiently differentiated in qualities. Quality guidance can lower the price of each product and benefit consumers while raising the average transaction price. With heterogeneous match distributions, a reward for the best initial inspection can help consumers with advertising fixed and harm them when advertising adjusts. A two-type design extension shows that no finite optimum over implementable outcomes uses positive weights on both quality and low prices, even when the platform values consumer surplus.
\end{abstract}
\medskip
\noindent\textit{Keywords}: consumer search, recommendation, firm heterogeneity, advertising, visiting rates.\\
\textit{JEL codes}: D43, D83, L13, L15.
\clearpage
\section{Introduction}

Online marketplaces use recommendation rules to reward low prices and good service. Amazon identifies price, delivery speed, and post-purchase satisfaction among its product-search signals. In October 2024, Etsy began giving search priority to eligible U.S. domestic listings with shipping charges below six dollars.\footnote{Amazon, ``How Customers Get to Pick the Winners in Amazon's Store,'' \href{https://www.aboutamazon.com/news/how-amazon-works/how-customers-get-to-pick-the-winners-in-amazons-store}{Amazon News}; Etsy, ``We're Updating How Shipping Price Is Factored Into Search for US Domestic Listings,'' October 3, 2024, \href{https://www.etsy.com/seller-handbook/article/1293023712519}{Seller Handbook}. The Etsy change took effect on October 1 and applied to eligible listings. Both sources accessed November 3, 2026.} During Pinduoduo's 2024 midyear sales campaign, a reported rule change gave discounted products greater weight in search and recommendation.\footnote{Zhou Wenjun, ``Pinduoduo: Price Cuts Earn Traffic during the June 18 Promotion,'' \emph{Dianshangbao}, republished by \href{https://column.iresearch.cn/b/202405/981964.shtml}{iResearch}, May 27, 2024. The report describes the promotional rule.} These decisions change a seller's return to cutting its price. They also change the competing products a buyer is likely to examine.

Sellers can influence those visits through advertising. Alibaba's paid-search ranking combines keyword bids with product or merchant quality, popularity, and customer feedback. Amazon's Sponsored Products lets sellers choose bids and budgets and pay for clicks.\footnote{Alibaba Group, \emph{Fiscal Year 2025 Annual Report}, p.~11, \href{https://www.hkexnews.hk/listedco/listconews/sehk/2025/0626/2025062601064.pdf}{annual report}. \label{fn:amazon-ads}Amazon Ads, ``Sponsored Products,'' \href{https://advertising.amazon.com/en-us/solutions/products/sponsored-products}{product documentation}, accessed November 3, 2026.} A recommendation reform---a change in the rule assigning exposure---therefore changes both pricing incentives and the return to advertising. Buyers, in turn, decide whether to accept the product they have found or incur the cost of examining another one. Does a stronger reward for low prices make this market more competitive? Does directing visits toward better products necessarily help consumers after sellers respond?

These questions also arise in policy discussions. China's 2026 quality-oriented e-commerce initiatives call for greater weight on product quality and service reputation in traffic allocation and for changes to display, recommendation, and marketing mechanisms.\footnote{Ministry of Commerce and five other agencies, ``Guiding Opinions on Better Serving the Real Economy and Promoting High-Quality E-Commerce Development,'' March 20, 2026, posted April 6, Part III(5), \href{https://www.mofcom.gov.cn/zcfb/zhzc/art/2026/art_7e1dc12697b744fd89e6110673ac117c.html}{official text}; Ministry of Commerce and six other agencies, ``Notice on Implementing the Five-Excellence Action for Quality E-Commerce,'' September 11, 2026, posted October 10, Part II(2), \href{https://www.mofcom.gov.cn/gztz/art/2026/art_9c50a751387d4ce7ab1f86cd9a949ccb.html}{official text}. Titles are author translations.} The European Union's Digital Services Act requires platforms to explain their main recommendation parameters and the reasons for their relative importance.\footnote{Regulation (EU) 2022/2065, Article 27, \href{https://eur-lex.europa.eu/eli/reg/2022/2065/oj/eng}{official text}.} Evaluating such rules requires understanding how their incentives affect the alternatives consumers can find and the purchases they make.

The difficulty is that sellers' choices and the value of continued search determine one another. A price cut makes a product more attractive and may earn it more visits. Advertising also changes the products buyers encounter. As these opportunities change, buyers adjust how much utility they require before stopping, which changes demand for every seller. Heterogeneity makes the adjustment difficult to infer from the directly rewarded product. A product suitable for many buyers may offer only a modest improvement when accepted; another may offer a large improvement to the few buyers it suits. Similar expected gains from inspection can therefore coexist with different sales probabilities, markups, and incentives to advertise.

We develop a framework for comparing recommendation rules in such markets. There is a continuum of sellers within each of finitely many types, which may differ in quality, production cost, match distribution, and advertising technology. The platform commits to a score based on public characteristics, prices, and advertising; relative scores determine which product is inspected next. Sellers choose advertising and then prices. Consumers learn their individual match upon inspection and decide whether to continue. The continuum represents a large assortment in which each seller has a negligible effect on aggregate opportunities, although a change in the rule can alter those opportunities through the responses of all sellers.

The equilibrium construction starts from a proposed reservation utility, the utility a consumer requires to stop searching. When advertising multiplies the price and characteristic component of the score, it scales operating profit at every price by the same factor. Each type's conditional price can therefore be found before solving its advertising choice. A common demand scale then reconciles advertising with aggregate sales. Finally, the products and visits generated by those choices must make the proposed stopping rule optimal. We establish existence, allowing consumer exit and zero advertising, and give a sufficient condition for uniqueness in terms of the dispersion of search opportunities and bounds on sellers' responses.

The common variables have distinct economic roles. Reservation utility determines how attractive a product must be to secure a purchase; the demand scale determines the sales return to its score. Adding seller types changes the individual calculations without increasing the number of common variables. This permits comparisons of recommendation rules that retain heterogeneous prices and advertising choices rather than impose a representative seller or a fixed distribution of visits. The effect of a rule separates into changes in the gain from inspecting each product and changes in the probabilities of encountering those products. Their sum determines consumers' willingness to continue and the ensuing price response. Acceptance probabilities then turn visits into transaction shares, allowing the same analysis to distinguish individual price changes from changes in what consumers buy.

Different score specifications recover uniform random search, fixed differences in exposure, price-dependent recommendation, and advertising for attention. The applications use this flexibility to answer three related questions. Does rewarding price cuts preserve the alternatives that sustain competition? Does a lower price at every seller imply a lower average price paid? Does rewarding a valuable inspection still help consumers when sellers compete for exposure? Each comparison uses the same conditional choices and search equation, but changes a different part of the economic adjustment.

The price application shows that a stronger reward lowers prices when product opportunities are identical. With different products, the reward may shift visits away from a more expensive product that offers better service or a better chance of a valuable match. Consumers become less willing to continue searching, making a sale less sensitive to price. This loss of competition can outweigh every seller's stronger direct incentive to cut price, so all equilibrium prices rise. A primitive sufficient condition establishes the reversal for general match distributions. A logistic benchmark gives an explicit classification, and strict reversals persist with positive quality rewards and small positive advertising expenditure and responses.

Quality guidance works through a different part of the same construction. At a given reservation utility it leaves conditional prices unchanged and favors better search opportunities. With common costs and match distributions, the improvement survives advertising adjustment even when advertising technologies differ. Consumers become more selective and, under uniqueness, prices weakly fall. Yet purchases may shift sufficiently toward the better, still more expensive product that the average transaction price rises while every seller cuts its price and consumers gain. An increase in the average price paid can therefore accompany stronger price competition. Marketplace data must distinguish changes in individual prices from changes in purchase shares.

With different match distributions, the chance of acceptance and the gain from inspection need not move together. A reward for the type initially offering the greatest expected gain can then help consumers with advertising fixed and harm them when advertising adjusts, with prices and search endogenous in both comparisons. In the construction, the favored type is likely to be accepted; another advertiser offers nearly as valuable but less frequent gains. Its advertising reduction allows a worse type with baseline exposure to gain visits. Consumers lose valuable alternatives, become less selective, and face higher prices. We give the response criterion, primitive conditions supporting a strict reversal, and cases in which the reversal cannot occur.

An extension lets the platform choose the price and quality weights, receiving advertising payments and a fixed commission and possibly valuing consumer surplus. In this two-type design benchmark, a rule with both weights strictly positive admits a nearby adjustment that preserves consumer surplus and raises receipts. Weakening the price reward raises sellers' prices; adjusting quality guidance to preserve the value of search shifts purchases toward the better product. These changes increase revenue without reducing consumer surplus. Hence no finite optimum over implementable outcomes uses both weights; the rule-choice implication also holds with locally unique continuation. The result uses the same conditional pricing, search consistency, and purchase composition that determine the recommendation comparisons.

The paper builds on the role of unexamined alternatives in price competition \citep{Wolinsky1986,AndersonRenault1999}. Prominence and ordered search make the identity of those alternatives important \citep{ArmstrongVickersZhou2009,Zhou2011}. \citet{ChoiDaiKim2018} show how optimal sequential search with partial product information and advertised prices can be represented as a discrete-choice pricing problem. Our representation addresses prices and advertising when a platform's score determines encounter probabilities. A seller takes the aggregate search environment as given, but the collective response to a rule changes that environment. The conditional price and scale calculations allow this response to be found without replacing heterogeneous sellers by a representative firm. Search intermediation and random quality offer related approaches to tractability \citep{ChenZhang2018,Chen2024}.

\citet{AndersonRenault2026} jointly analyze heterogeneous demand, pricing, and position bids. Their position externalities link the products encountered earlier or later to demand and the willingness to continue search, and support a characterization of preferred rankings. We study the distribution of encounters generated by a score that responds to sellers' actions. This permits the same equilibrium construction to evaluate price rewards and rewards for public quality or match characteristics, retaining advertising and the composition of completed purchases. Advertising for attention is endogenous in \citet{HaanMoraga2011}; paid placement and position auctions link advertising returns to search and relevance \citep{AtheyEllison2011,ChenHe2011}. Here differences in advertising responses also determine whether a recommended improvement survives sellers' competition for exposure.

Information and search-pool design can alter both allocation and price competition \citep{ArmstrongZhou2022,EliazSpiegler2016}, a trade-off also central to empirical platform design \citep{DinersteinEtAl2018}. \citet{Zhou2022} shows that improved information or search-pool selection can shorten search, raise prices, and harm consumers. Our fixed-versus-adjusting-advertising comparison isolates a different change: the same direct recommendation reward benefits consumers with endogenous prices and search when advertising is fixed, and loses that benefit when advertising adjusts. Prominence payments and targeting provide further ways to influence the products buyers encounter \citep{ArmstrongZhou2011,deCorniere2016}.

Recent platform-design work studies visit-allocation contracts with privately heterogeneous sellers \citep{Zhang2026}, informative recommendations \citep{PeitzSobolev2026a,PeitzSobolev2026b}, sponsored and organic positions \citep{JanssenEtAl2026}, and information design with advertising competition \citep{KeLinLu2026}. Endogenous prices also affect optimal search precision and match recommendations \citep{Zhong2023,Shi2025}. Our rule specifies how public characteristics and seller choices affect exposure, so comparing rules requires following the induced pricing, advertising, and search responses together. Personalized ranking can weaken the direct return to a price cut \citep{QiuEtAl2025}, and automated bidding changes prices and the return to sponsored exposure \citep{SaremiSubramanian2026}. In our price reversal, the direct return to cutting price strengthens; it is the loss of valuable alternatives that allows equilibrium prices to rise.

Section~\ref{sec:model} presents the market and derives demand. Section~\ref{sec:core} constructs equilibrium and derives the effects of changes in recommendation. Section~\ref{sec:applications} applies these results to price rewards, quality guidance, and rewards across match distributions. Section~\ref{sec:extensions} studies recommendation design, joint pricing--advertising scores, and the demand approximation under optimal search in a finite assortment. The appendices provide proofs and verification of the numerical illustration.

\section{Recommendation and Consumer Search}\label{sec:model}
\subsection{Products, consumers, and recommendation}

There is a unit mass of firms and a unit mass of consumers. Firms belong to one of $n$ publicly observed types. Type $i$ has mass $\alpha_i>0$, with $\sum_i\alpha_i=1$, and contains a continuum of firms. Its characteristics include product quality $v_i$, marginal cost $c_i\geq0$, and the technology through which advertising affects recommendation. A firm chooses its advertising expenditure $z_i\geq0$ and its price $p_i\geq c_i$.

A consumer who buys a type-$i$ product at price $p$ obtains utility $v_i-p+\varepsilon_i$. An offer consists of a product at its posted price. The match value $\varepsilon_i$ has distribution $F_i$ and is independent across firms and consumers. Consumers know the aggregate distribution of public characteristics and posted prices, but learn their match with a particular product only by inspecting it. Inspection is also required to purchase. Previous offers can be recalled, and a consumer can leave with payoff zero.

Public quality is valued equally by all buyers; individual fit varies across them. This distinction also underlies the product-design analysis of \citet{BarIsaacEtAl2012}. For example, appliances may have publicly observed service packages, while buyers must inspect their dimensions and features to learn whether they suit the intended use. Recommendation can favor an observable service attribute even though inspection remains necessary. Production costs separately determine the seller's return from a sale.

Consumers choose whether to inspect another product; the platform's rule determines the distribution of the next inspection. The rule is common across consumers and remains stationary during a search. This describes discovery through a stream of recommended offers in a large assortment, where examining a few products has little effect on the distribution still available. Section~\ref{sec:extensions} gives a finite-assortment foundation for this approximation.

Advertising changes the chance of receiving a visit while leaving product quality and the information learned at inspection unchanged. Initially, advertising expenditure is a real resource cost, such as staff time used to manage a campaign. Expenditure units can also represent promotional effort with a strictly increasing cost, by incorporating that cost into the advertising technology. Section~\ref{sec:applications} distinguishes resource costs from payments retained by the platform.

A \emph{visiting rate} is the probability density of the next inspection per unit mass of firms. Let $\rho_i$ be the visiting rate for a firm of type $i$. The \emph{visit share} of that type is $\alpha_i\rho_i$: the probability that the next inspected firm belongs to it. At a type-symmetric action profile,
\begin{equation}
 \sum_i\alpha_i\rho_i=1.
 \label{eq:visiting-rates}
\end{equation}
Uniform random search has $\rho_i=1$, so each type's visit share equals its population share $\alpha_i$. Recommendation changes visiting rates in response to product characteristics and firms' choices. For a type-specific quantity $Y_i$, write $\E_{\rho}[Y_i]=\sum_i\alpha_i\rho_iY_i$ for its average over the next inspection. Covariances denoted by $\Cov_{\rho}$ use the same type probabilities $\alpha_i\rho_i$.

The platform implements the rule through positive \emph{recommendation scores}. A type-$i$ firm charging $p$ and spending $z$ receives score $w_i(p,z)>0$. At a type-symmetric profile, write $w_i=w_i(p_i,z_i)$. Visiting rates are
\begin{equation}
 \rho_i=\frac{w_i}{\sum_j\alpha_jw_j}.
 \label{eq:score-rate}
\end{equation}
Only relative scores matter: multiplying every score by the same positive number changes no visiting rate. A single infinitesimal firm takes the aggregate score as given, so increasing its score raises its visiting rate proportionally. When all firms respond, both their scores and the aggregate score can change.

Equal constant scores give uniform random search; unequal constant scores give persistent differences in exposure. A score that decreases with price makes cheaper offers more likely to be inspected. An advertising multiplier allows firms to compete for these visits. Comparisons without advertising remove that choice.

The platform commits to the rule first. Firms choose advertising expenditure, observe the aggregate advertising expenditure history, and then choose prices. Consumers search last. A firm's price choice can therefore respond to its advertising expenditure, while consumers' search decisions respond to the resulting prices and visits.

We study pure firm-strategy equilibria: firms choose advertising and prices without randomizing. After advertising is chosen, histories with the same type-specific advertising expenditure distributions lead to the same aggregate outcome in the subsequent pricing and consumer-search stages. We call this an \emph{aggregate-invariant continuation selection}. A single infinitesimal firm's advertising deviation therefore leaves those aggregate outcomes unchanged, while the firm remains free to choose its own subsequent price.

Consumer tie-breaking is stationary and depends on aggregate offers. Feasible action sets are compact and sufficiently wide that the bounds never constrain an optimal choice. Appendix~\ref{app:sequential} gives the formal implementation, including histories with different advertising expenditures among firms of the same type.

\subsection{Consumer search and firm demand}

The first inspection is free; each additional inspection costs $s>0$. This includes the time needed to evaluate a product and compare it with offers already examined. Let $f_i$ be the density of $F_i$. Write $S_i(x)=1-F_i(x)$ for the probability that a match exceeds the required value $x$. The hazard rate $h_i$ measures how fast this probability falls as the requirement rises, and $G_i$ gives the expected gain above the requirement:
\begin{equation}
 h_i(x)=\frac{f_i(x)}{S_i(x)},\qquad
 G_i(x)=\int_x^\infty S_i(y)\,\dd y.
 \label{eq:distribution-objects}
\end{equation}
The expected gain accounts for both the chance of acceptance and how far the match exceeds the requirement: $G_i(x)/S_i(x)$ is the expected excess match conditional on acceptance. Two products can therefore offer similar inspection gains even when one is much more likely to be accepted. This distinction will matter when recommendation rewards products with different match distributions.

\begin{assumption}[Match distributions]\label{ass:match-en}
Each $F_i$ has full real support, a positive continuous density, a finite expectation of the match value truncated below at zero, and a nondecreasing hazard rate. Results using derivatives impose the additional smoothness they require.
\end{assumption}

A nondecreasing hazard makes the proportional loss of purchases from a price increase weakly larger when the product is already difficult to accept. This property gives the conditional pricing problem its monotonicity. Full support allows matches above and below every finite cutoff, and the tail condition makes the gain from another inspection finite.

Let $u\geq0$ be the consumer's \emph{reservation utility}, the utility required to accept an offer instead of continuing to search. An inspected type-$i$ product is accepted when $v_i-p_i+\varepsilon_i\geq u$. Its match value must therefore exceed $u+p_i-v_i$. The acceptance probability is $1-F_i(u+p_i-v_i)$, and the expected gain above the reservation utility is $G_i(u+p_i-v_i)$. Optimal search requires
\begin{equation}
 \E_{\rho}G_i(u+p_i-v_i)=s.
 \label{eq:consumer-search}
\end{equation}
Note that recall does not change this rule: previously rejected offers lie below the same requirement. $u>0$ corresponds to the case \emph{active search}, while $u=0$ means further search is never worthwhile, so that the consumer stops after the free inspection. Appendix~\ref{app:basic} also treats indifference and exit at this boundary.

Under active search, the probability of purchasing on any inspection is the visit-share average $\E_{\rho}[1-F_i(u+p_i-v_i)]$. Its reciprocal is the expected number of inspections each consumer makes before buying. A firm's demand is its visiting rate times the probability of purchase conditional on a visit, multiplied by this total number of inspections:
\begin{align}
 D_i&=\frac{\rho_i[1-F_i(u+p_i-v_i)]}{\E_{\rho}[1-F_i(u+p_i-v_i)]}
     =\frac{w_i[1-F_i(u+p_i-v_i)]}{C},\notag\\
 C&=\sum_i\alpha_iw_i[1-F_i(u+p_i-v_i)].
 \label{eq:demand-en}
\end{align}
Here $D_i$ is demand per unit mass of type-$i$ firms, and $C$ is the \emph{common demand scale}, the denominator that makes aggregate sales equal the unit mass of consumers. Given the market aggregates, a firm chooses advertising and then price to maximize
\[
 (p-c_i)D_i(p,z)-z.
\]
A higher price increases the markup on each sale but can reduce both visits and the chance of acceptance after a visit. Advertising can increase visits, at a cost.

A single firm takes the demand scale as given. For its given score and acceptance probability, a higher $C$ means fewer sales and a lower return to advertising. Collective changes in scores and acceptance determine that scale in equilibrium. Rescaling every score rescales $C$ by the same factor and leaves demand unchanged.

\subsection{Transactions, profit, and consumer surplus}

Visit shares and transaction shares need not coincide. A product may receive many visits but few purchases if its price is high relative to its value. Define the \emph{transaction share} of type $i$ as
\[
 \omega_i=\frac{\alpha_i\rho_i[1-F_i(u+p_i-v_i)]}{\E_{\rho}[1-F_i(u+p_i-v_i)]}.
\]
These shares sum to one under active search, because every consumer eventually buys. Averages denoted by $\E_\omega$ use transaction shares. Visit shares describe the next opportunity a consumer expects to inspect. Transaction shares describe what consumers ultimately buy and hence the sales on which firms earn their markups.

Let $\CS$ denote consumer surplus, $\Ads$ aggregate advertising expenditure, $\Pi$ industry net profit, and $W$ total surplus. Then
\begin{equation}
 \CS=u+s,\qquad \Ads=\sum_i\alpha_i z_i,\qquad
 \Pi=\E_\omega[p_i-c_i]-\Ads,\qquad W=\CS+\Pi.
 \label{eq:welfare}
\end{equation}
The markup on a type-$i$ sale is $p_i-c_i$. The free first inspection makes consumer surplus exceed the value of entering a paid search by $s$. Total purchases have unit mass, so industry operating profit is the transaction-weighted markup, $\E_\omega[p_i-c_i]$. Net profit deducts advertising expenditure. Since advertising is treated as a real resource cost, total surplus makes the same deduction. These formulas apply to active search; the appendix gives the accounting when some consumers leave without buying.

\section{Equilibrium and the Feedback from Consumer Search}\label{sec:core}
\subsection{Firm choices and equilibrium}

We now determine prices, advertising, and consumers' stopping decisions jointly. The feedback runs through continued search: firms change the opportunities consumers expect to find, and consumers' willingness to continue changes firms' returns. The main analysis uses scores in which advertising multiplies the price component:
\begin{equation}
 w_i(p,z)=w_i(p,0)g_i(z),\qquad g_i(0)=1.
 \label{eq:separable}
\end{equation}
The zero-advertising score $w_i(p,0)$ describes how price and the firm's public characteristics enter recommendation; $g_i$ is the \emph{score multiplier} generated by advertising. Thus advertising changes the number of visits a firm can obtain without changing the proportional reward to a price cut. The normalization $g_i(0)=1$ leaves scores unchanged if an original positive baseline multiplier is incorporated into $w_i(p,0)$. Section~\ref{sec:extensions} examines rules in which advertising also changes pricing incentives.

\begin{assumption}[Recommendation scores and advertising]\label{ass:scores-en}
Each zero-advertising score $w_i(p,0)$ is positive, continuously differentiable, nonincreasing, and log-concave in price. Each $g_i$ is continuous on $[0,\infty)$, continuously differentiable on $(0,\infty)$, strictly increasing, and strictly concave, with $g_i(0)=1$ and $g_i'(z)\to0$ as $z\to\infty$. Its right derivative at zero may be finite or infinite.
\end{assumption}

A price increase weakly reduces the firm's visiting rate in proportion to the decrease in its score. Log-concavity makes that proportional loss nondecreasing in price. Concavity of $g_i$ gives diminishing returns to advertising and allows zero expenditure when the initial return is too low.

For a given reservation utility, a seller faces two sources of demand loss from charging more: fewer visits and a lower probability of acceptance after each visit. Separability makes advertising multiply demand at every price by the same factor. The conditional pricing problem can therefore be solved before advertising and market consistency are considered.

\begin{lemma}[Conditional prices]\label{lem:conditional-price}
Maintain Assumptions~\ref{ass:match-en}--\ref{ass:scores-en}. At every proposed reservation utility $u>0$, type $i$ has a unique globally optimal price $p_i(u)$, independent of advertising and the common demand scale. It satisfies
\begin{equation}
 \frac{1}{p_i-c_i}=h_i(u+p_i-v_i)
 -\left.\frac{\partial\log w_i(p,0)}{\partial p}\right|_{p=p_i}.
 \label{eq:conditional-price}
\end{equation}
If hazards are continuously differentiable and price scores are twice continuously differentiable, then
\begin{equation}
 -1<\frac{\partial p_i}{\partial u}\leq0.
 \label{eq:price-response}
\end{equation}
\end{lemma}

The left side of~\eqref{eq:conditional-price} is the proportional gain in the markup from a price increase. The right side adds the proportional losses in acceptance and visits. The markup gain falls with price, whereas both demand losses weakly rise, yielding a unique optimum. More demanding consumers weakly lower the seller's best price, but any price cut is smaller than the increase in reservation utility. The required match therefore rises: price adjustment moderates the increase in selectivity without offsetting it. These conditional prices also apply after any advertising history, since advertising expenditure is then sunk and multiplies the price objective.

It remains to reconcile the sellers' returns to advertising with aggregate sales and consumers' search decisions. At the conditional prices, the advertising first-order condition is
\begin{equation}
 \frac{(p_i-c_i)w_i(p_i,0)[1-F_i(u+p_i-v_i)]}{C}\,g_i'(z_i)\leq1,
 \quad\text{with equality if }z_i>0.
 \label{eq:advertising-choice-main}
\end{equation}
The derivative at zero is the right derivative. Positive advertising equates its marginal operating return to its unit marginal cost. A larger common scale $C$ reduces sales at every given score, so each seller spends weakly less. This gives a particularly simple way to clear the conditional market.

\begin{theorem}[Market equilibrium]\label{thm:aggregation-en}
Maintain Assumptions~\ref{ass:match-en}--\ref{ass:scores-en} and use the conditional prices in Lemma~\ref{lem:conditional-price}.
\begin{enumerate}
\item For every $u>0$, there is a unique positive scale $C(u)$ and a unique advertising expenditure $z_i(u)$ for each type satisfying~\eqref{eq:advertising-choice-main} and
\begin{equation}
 C=\sum_i\alpha_iw_i(p_i,0)[1-F_i(u+p_i-v_i)]g_i(z_i).
 \label{eq:scale-main}
\end{equation}
\item Let $\rho_i(u)$ be the resulting visiting rates and define the conditional gain from another inspection by
\begin{equation}
 M(u)=\sum_i\alpha_i\rho_i(u)G_i\bigl(u+p_i(u)-v_i\bigr).
 \label{eq:conditional-search-gain}
\end{equation}
Active-search equilibria are exactly the solutions of $M(u)=s$ with $u>0$. An equilibrium always exists, allowing a zero reservation utility. Every consistent outcome is implemented as a subgame-perfect equilibrium within the aggregate-invariant continuation class.
\end{enumerate}
\end{theorem}

The scale equation has a unique solution because its left side rises with $C$, while its right side falls weakly as sellers reduce advertising. This argument permits zero advertising and different technologies across types. The remaining equation asks whether the inspection opportunities generated by these choices make consumers willing to use the proposed stopping rule. At the participation boundary, the probability of continuing after an unsuccessful inspection replaces the positive reservation utility as the consumer variable; Appendix~\ref{app:basic} establishes existence across both branches.

The two common variables have distinct economic roles. Reservation utility determines the acceptance requirement faced by each seller; the demand scale determines the sales return to its score. Adding types adds individual choice problems but leaves these two consistency requirements unchanged. Thus the same construction can compare recommendation rules while retaining differences in quality, costs, matches, and advertising technologies. A product may offer a large gain from inspection yet give its seller little reason to advertise because acceptance is unlikely or its markup is small. Solving the choices separately and then imposing search consistency preserves precisely these differences.

Conditional uniqueness does not by itself imply a unique market equilibrium: different reservation utilities may be supported by the prices and advertising they induce. The next step identifies how these responses change the value of continued search.

\subsection{How firms' responses change the value of search}

As reservation utility rises, the conditional market in Theorem~\ref{thm:aggregation-en} changes in two ways. Each product offers less gain above the new requirement, and sellers' price and advertising responses redistribute visits. Lemma~\ref{lem:conditional-price} establishes the first effect: the required match rises by $1+\partial p_i/\partial u>0$, even after the seller cuts its price. The second effect depends on which products become easier to encounter.

Different price and advertising responses also change visit shares. Combining the changes within each type with this redistribution gives
\begin{align}
 \frac{\partial M}{\partial u}
 ={}&-\underbrace{\E_{\rho}\!\left[(1-F_i(u+p_i-v_i))
                  \left(1+\frac{\partial p_i}{\partial u}\right)\right]}_{\text{lower gain from each type}}\notag\\
 &+\underbrace{\Cov_{\rho}\!\left(G_i(u+p_i-v_i),
                         \frac{\partial\log w_i}{\partial u}\right)}_{\text{change in visit shares}}.
 \label{eq:feedback}
\end{align}
The first term in $\partial M/\partial u$ collects the decline in each type's gain. The covariance captures the change in visit shares. It is taken across types using visit-share weights, with $\partial\log w_i/\partial u$ measuring the proportional score change after conditional prices, advertising, and the scale have adjusted.

The covariance is positive when visits move toward products offering larger gains, and negative when visits move toward smaller gains. It values the change in the mixture of inspections. Only relative score responses matter: if every score rises proportionally, visit shares stay unchanged and the term is zero. It is also zero when all products offer the same gain.

A seller's price and advertising choices thus affect its rivals by changing what consumers expect to find if they continue searching. Gaining visits can make consumers more demanding when the product is a valuable alternative, or less demanding when it displaces better alternatives. Public quality alone need not rank these opportunities: prices and match distributions also matter. Equation~\eqref{eq:feedback} separates their search value from the sellers' responses that redistribute visits.

One use of the decomposition is to establish uniqueness. If the redistribution of visits cannot offset the decline in gains within each type, $M$ is decreasing and the search equation has at most one solution. The test can be checked over a candidate interval before solving the market. For this test, suppose each type on its own supports active search. Let $u_i^s>0$ be the reservation utility that makes consumers indifferent about another inspection if every inspected product belongs to type $i$, with its price chosen conditionally. Thus $G_i(u_i^s+p_i(u_i^s)-v_i)=s$. The superscript $s$ denotes indifference at the given search cost. Define
\begin{equation}
 I=\left[\min_i u_i^s,\ \max_i u_i^s\right].
 \label{eq:root-interval}
\end{equation}
Every active equilibrium lies in $I$: below it every type offers a gain exceeding $s$, and above it every type offers less. Its positive lower endpoint also excludes the participation boundary.

To state a test that does not require solving for market advertising, write $x_i(u)=u+p_i(u)-v_i$ for the conditional match cutoff and let
$b_i=-\partial_p\log w_i(p,0)|_{p=p_i}$ be the proportional loss of score from raising price. In the following expressions, $h_i$, $S_i$, and $G_i$ are evaluated at $x_i(u)$. Let $\bar\chi$ bound the advertising multiplier's response to its operating return:
\[
 \bar\chi=\max_i\sup_{0<z\leq\bar z_i}
       \frac{-g_i'(z)^2}{g_i(z)g_i''(z)}<\infty.
\]
Here $\bar z_i$ is an upper bound on conditional advertising obtained directly from the primitives.\footnote{At the conditional prices, set $C^{\rm org}(u)=\sum_j\alpha_jw_j(p_j,0)S_j$. Let $\bar z_i$ maximize $g_i(z)\max_{u\in I}\{(p_i-c_i)w_i(p_i,0)S_i/C^{\rm org}(u)\}-z$. Since $C(u)\geq C^{\rm org}(u)$, no conditional advertising choice exceeds $\bar z_i$. An empty positive-advertising interval contributes zero to $\bar\chi$.} At an interior advertising choice, a one-percent increase in the operating return per unit of multiplier raises the optimal multiplier by at most $\bar\chi$ percent. This bound depends on the shape of the advertising technology, rather than on the size of its expenditure alone.

Define $\Delta_G(u)=\max_iG_i-\min_iG_i$, the range of inspection gains. The following quantity bounds the range of proportional score responses:
\begin{align}
 \Delta_J(u)={}&\max_i\{-b_i\partial_u p_i\}
                   -\min_i\{-b_i\partial_u p_i\}\notag\\
 &+\bar\chi\left[\max_i h_i
             -\min_i\{h_i+(h_i+b_i)\partial_u p_i\}\right].
 \label{eq:visibility-response-bound}
\end{align}
The first line measures differences in score changes due to repricing. The second bounds the additional differences caused by advertising, including the common change in its return. All terms can be computed from conditional prices and primitives. Their product with $\Delta_G$ bounds how much visit reallocation can offset the declining gain from each type.

\begin{proposition}[A sufficient condition for unique equilibrium]\label{thm:uniqueness}
Maintain Assumptions~\ref{ass:match-en}--\ref{ass:scores-en}. Suppose every $u_i^s$ is strictly positive, hazards are continuously differentiable, and price scores are twice continuously differentiable on the conditional domains generated by $I$. Require $g_i\in C^2$ and $g_i''<0$ on each nonempty interval $(0,\bar z_i]$, with $\bar\chi<\infty$. If, for every $u\in I$,
\begin{equation}
 \Delta_G(u)\Delta_J(u)
 <4\min_i\{S_i(x_i(u))[1+\partial_u p_i(u)]\},
 \label{eq:primitive-unique}
\end{equation}
the pure firm-strategy equilibrium outcome is unique and has active search. Zero advertising is allowed. Every allowed advertising history also has a unique price and search continuation.
\end{proposition}

The right side is four times the smallest decline in a type's inspection gain. The left side bounds the reallocation effect through the dispersion of gains and responses. The factor four comes from the covariance bound for two bounded variables. Under~\eqref{eq:primitive-unique}, this reallocation cannot offset the decline within products, so $M$ is strictly decreasing. Appendix~\ref{app:uniqueness} proves that the argument continues to hold when firms enter or leave advertising.

Two useful cases follow. If the price-induced reallocation leaves a uniformly positive margin in~\eqref{eq:primitive-unique}, sufficiently small advertising responses preserve uniqueness; with price-independent scores, that margin is automatically positive. Alternatively, sufficiently similar inspection gains ensure uniqueness when the acceptance terms stay bounded away from zero and price and advertising responses remain bounded. Both cases allow positive advertising. Applied throughout a policy interval, the test makes the comparison follow a unique market outcome.

\subsection{Comparative statics}

To evaluate a recommendation reform, first solve the conditional market at the initial reservation utility. Sellers may cut prices to obtain visits or change advertising as its return changes. The resulting opportunities determine how consumers adjust their stopping decisions. Let $\eta$ index an exogenous change in the recommendation rule, keeping search cost, qualities, production costs, match distributions, and advertising technologies fixed. In this subsection write $p_i(u;\eta)$ and $M(u;\eta)$ to show that dependence. A partial derivative with respect to $\eta$ holds $u$ fixed and allows all conditional firm choices to adjust.

The conditional effect on search gains is $\partial M/\partial\eta$. Its sign determines whether consumers become more or less selective; the slope $\partial M/\partial u$ determines the adjustment needed to restore search indifference.

\begin{lemma}[Local equilibrium responses]\label{thm:policy}
Consider an active equilibrium with $\partial M/\partial u<0$ and a locally fixed set of firms choosing positive advertising. Assume the continuous price, policy, and advertising expenditure derivatives used in Appendix~\ref{app:policy-proof} exist, with strictly negative advertising curvature at positive choices. Along the locally unique differentiable path of equilibria, a rule change at fixed search cost satisfies
\begin{align}
 \frac{\dd u}{\dd\eta}
 &=-\frac{\partial M/\partial\eta}{\partial M/\partial u},
 &\frac{\dd\CS}{\dd\eta}&=\frac{\dd u}{\dd\eta},\notag\\
 \frac{\dd p_i}{\dd\eta}
 &=\frac{\partial p_i}{\partial\eta}
   +\frac{\partial p_i}{\partial u}\frac{\dd u}{\dd\eta}.
 \label{eq:policy-response-core}
\end{align}
The first price term is the seller's response at the initial reservation utility; the second is its response to the change in consumers' stopping decisions. When a type enters or leaves positive advertising and the stated derivatives exist on each side, the identities apply with the corresponding one-sided derivatives.
\end{lemma}

To find the reform's conditional effect on search gains, use the same distinction between changes within products and changes in visit shares:
\begin{align}
 \frac{\partial M}{\partial\eta}
 ={}&-\E_{\rho}\!\left[(1-F_i(u+p_i-v_i))
                         \frac{\partial p_i}{\partial\eta}\right]\notag\\
 &+\Cov_{\rho}\!\left(G_i(u+p_i-v_i),
                            \frac{\partial\log w_i}{\partial\eta}\right).
 \label{eq:policy-gain-main}
\end{align}
The score response includes conditional price and advertising adjustments at the given $u$, with the demand scale also adjusted. A price cut raises the gain from inspecting that product; the first term adds these improvements using visit shares. The second records whether visits move toward products providing higher or lower gains. A reform can lower every conditional price yet reduce the average search gain if it moves enough visits toward worse alternatives.

The direct score reward need not determine the final direction of visits. Advertising responds to the changed returns, which also depend on the common demand scale. A type choosing zero advertising retains a positive score because $g_i(0)=1$; it can gain visits when other firms reduce advertising expenditure, even without receiving a direct reward.

If the conditional effect is positive, continued search is worth more than its cost at the old reservation utility. Consumers raise their acceptance requirement until the expected gain again equals search cost. This higher requirement weakly lowers each conditional price. A negative conditional effect reverses these steps. The price equation in Lemma~\ref{thm:policy} adds this response to the reform's direct pricing effect. Thus a rule that rewards price cuts can nevertheless raise equilibrium prices if it sufficiently weakens the alternatives consumers expect to find. A smaller absolute slope of $M$ requires a larger reservation-utility adjustment to restore indifference.

The same reform also changes the composition of completed purchases. Let $\E_\omega[p_i]$ be the average transaction price, using the transaction shares defined in Section~\ref{sec:model}. Along any differentiable policy comparison,
\begin{equation}
 \frac{\dd}{\dd\eta}\E_\omega[p_i]
 =\underbrace{\E_\omega\!\left[\frac{\dd p_i}{\dd\eta}\right]}_{\text{individual price changes}}
 +\underbrace{\sum_i p_i\frac{\dd\omega_i}{\dd\eta}}_{\text{change in transaction composition}}.
 \label{eq:transaction-price-decomposition}
\end{equation}
Because transaction shares sum to one, the second term is positive when purchases move toward more expensive products. This change in what consumers buy can outweigh price cuts at every seller. With common cost and active search, the average transaction markup is also industry operating profit; net profit deducts advertising expenditure. Visits determine the alternatives consumers expect to inspect, whereas transactions determine the markups firms earn.

The applications now use these equations to compare three recommendation changes. Price rewards change both conditional prices and visit shares. Quality and match rewards leave conditional prices unchanged and act through visits and advertising. We begin with the price comparison: whether better incentives to cut prices also preserve the alternatives that discipline sellers.

\section{Recommendation, Transaction Selection, and Advertising}\label{sec:applications}

The following applications change the price reward, the quality reward, and the treatment of different match distributions. Each uses the conditional choices and search equation of Section~\ref{sec:core}. The comparisons differ in which part of the search-gain decomposition changes and in how the resulting visits translate into purchases.

\subsection{A common benchmark}\label{sec:common-benchmark}

The price and quality comparisons have common marginal cost $c$ and a common match distribution $F$, with hazard $h$ and expected search gain $G$ as defined in Section~\ref{sec:model}. Qualities and advertising technologies may differ. Section~\ref{sec:match-guidance} instead allows heterogeneous match distributions $F_i$, retaining common quality, cost, and advertising technology. The benchmark recommendation score is
\begin{equation}
 w_i(p,z)=\exp(kv_i-\beta p)g_i(z),\qquad k,\beta\geq0.
 \label{eq:applications-score}
\end{equation}
The quality weight $k$ rewards public quality; the price weight $\beta$ is the proportional score loss from a price increase. They can be varied independently. Each advertising multiplier $g_i$ satisfies Assumption~\ref{ass:scores-en}. A comparison without advertising sets the multiplier to one and removes the advertising expenditure choice.

Lemma~\ref{lem:conditional-price} gives the common conditional pricing equation
\begin{equation}
 p_i-c=\frac{1}{\beta+h(u+p_i-v_i)}.
 \label{px:main-price}
\end{equation}
Charging more loses demand in two ways: the price weight reduces the chance of a visit, and the acceptance hazard measures the loss of purchases after a visit. Quality guidance and advertising instead multiply the return at every price by the same factor. They leave conditional pricing unchanged but affect equilibrium prices through the products consumers expect to encounter and their willingness to continue searching.

\begin{table}[htbp]
\centering
\caption{Three recommendation comparisons in the same conditional market}
\label{tab:application-comparisons}
\small
\begin{tabular}{@{}>{\raggedright\arraybackslash}p{0.19\textwidth}>{\raggedright\arraybackslash}p{0.37\textwidth}>{\raggedright\arraybackslash}p{0.37\textwidth}@{}}
\toprule
Change & At a fixed reservation utility & Equilibrium adjustment \\
\midrule
Higher price weight $\beta$ & Each firm cuts its conditional price; visit shares and advertising also adjust. & The change in search opportunities can reinforce or overturn the direct price cuts. \\[5pt]
Higher quality weight $k$ & Conditional prices stay fixed; the quality component favors better products while advertising and visit shares adjust. & Better search opportunities weakly raise reservation utility and lower prices. \\[5pt]
Reward for the best inspection & Conditional prices stay fixed; a type-specific score multiplier changes advertising returns and visit shares. & With heterogeneous matches, advertising responses can preserve or reverse the improvement in search opportunities. \\
\bottomrule
\end{tabular}
\end{table}

Table~\ref{tab:application-comparisons} identifies the conditional change to be evaluated before the search equation determines the response of reservation utility.

\subsection{Price rewards and search competition}\label{sec:prices}

A reward for low prices gives sellers a reason to cut their prices to attract visits. It also makes relatively expensive products harder to find. When those products offer better service or reliability, their loss of exposure weakens the alternatives available to consumers who reject an offer. The framework allows us to compare the stronger incentive to cut a price with this change in the competition sustained by continued search.

For a differentiable hazard, write $h_i=h(u+p_i-v_i)$ and $h_i'=h'(u+p_i-v_i)$ for its value and slope at type $i$'s acceptance cutoff. Differentiating the conditional pricing equation~\eqref{px:main-price} along an equilibrium path gives
\begin{equation}
 \frac{\dd p_i}{\dd\beta}
 =-\frac{1+h_i'\,\dd u/\dd\beta}{(\beta+h_i)^2+h_i'}.
 \label{eq:application-price-response}
\end{equation}
The unit term in the numerator is the direct incentive: at a given reservation utility, every seller cuts its price. The other term follows from consumer search. If the resulting opportunities are better, consumers become more selective and sellers cut prices further. If valuable alternatives instead lose enough visits, consumers become less selective and sellers raise prices in response. This is the conditional-price and reservation-utility decomposition in Lemma~\ref{thm:policy}. It applies at any fixed quality weight and allows endogenous advertising.

With identical qualities, reallocating visits cannot remove better alternatives. Search indifference fixes the common match requirement $x_s$, defined by $G(x_s)=s$, and the active-equilibrium price is $c+1/[\beta+h(x_s)]$. A larger price weight lowers this price and raises consumer surplus, even if advertising technologies differ. To obtain a reversal, the products displaced must offer sufficiently different search opportunities.

\paragraph{A condition on product and match characteristics.}
The following construction gives such a condition for a general smooth match distribution satisfying Assumption~\ref{ass:match-en}. Choose two match cutoffs $x_H<x_L$ with positive hazard slopes, and a baseline visit share $q\in(0,1)$ for type $H$. In the next result, $h_i=h(x_i)$, $h_i'=h'(x_i)$, $S_i=1-F(x_i)$, and $G_i=G(x_i)$; $\E_q$ averages across the two types with weights $q$ and $1-q$. These distributional quantities determine a quality gap and search cost:
\begin{equation}
 v_H-v_L=\frac1{h_H}-\frac1{h_L}+x_L-x_H,
 \qquad s=qG_H+(1-q)G_L.
 \label{eq:px-general-primitive-class}
\end{equation}
For any fixed $k\geq0$, choose type masses satisfying
$\alpha_H/\alpha_L=[q/(1-q)]e^{-k(v_H-v_L)}$.
At $\beta=0$ and without advertising, the quality reward then gives visit share $q$. Set the common quality level so that $v_L-c>1/h(x_s)-x_s$. This makes search worthwhile even if every visit is to the low-quality type.

\begin{proposition}[Price rewards and the loss of search opportunities]\label{prop:px-general}
Consider the two-type market constructed above, initially without advertising. If
\begin{equation}
 q(1-q)(G_H-G_L)\left(\frac1{h_H}-\frac1{h_L}\right)
 >\E_q\frac{S_i}{h_i^2+h_i'}
   +\frac{\E_q[S_i h_i^2/(h_i^2+h_i')]}{\min\{h_H',h_L'\}},
 \label{eq:px-general-primitive-inequality}
\end{equation}
then there is a positive interval of price weights starting at zero on which equilibrium is unique and active. Increasing the price weight within that interval raises both prices and lowers consumer surplus. The strict comparisons persist under sufficiently small admissible changes in the distribution and market primitives. They also hold for a common strictly concave advertising technology under which both types advertise, with arbitrarily small expenditure and responses of their multipliers to proportional changes in operating returns.
\end{proposition}

The left side of~\eqref{eq:px-general-primitive-inequality} is the loss from moving visits toward the cheaper product. It is larger when the products differ more in the gains they offer and in their prices. The first term on the right is the improvement from conditional price cuts. The second makes the condition strong enough to raise both equilibrium prices: the deterioration in search opportunities must lower reservation utility by more than either seller's direct incentive can offset. All terms are evaluated at the chosen distributional cutoffs. They define a class of markets before a market equilibrium is solved.

The proof substitutes these quantities into the two terms of~\eqref{eq:policy-gain-main} and then uses the price response above. The same calculation therefore identifies both the source of the reversal and a sufficient condition for it. Appendix~\ref{px:general-distributions} verifies the construction and its neighborhood. The condition is satisfied by both logistic and normal matches.

\paragraph{An exact logistic comparison.}
For a sharper classification, take two equally numerous types and standard logistic matches, $F(x)=1/(1+e^{-x})$. Here $h(x)=F(x)$ is the rejection probability and $G(x)=-\log h(x)$. Retain the entry condition $v_L-c>1/h(x_s)-x_s$. Begin without advertising and with $k=0$, so an initially price-neutral rule gives equal visit shares. The higher-quality seller charges more but offers a higher acceptance probability and a larger search gain. A new price reward consequently improves each conditional price while shifting visits away from the better opportunity.

\begin{proposition}[The logistic benchmark]\label{px:proposition}\label{prop:px-positive-advertising}
Consider the two-type logistic market with equal masses and the entry condition above. Parts 1 and 2 have no advertising and $k=0$.
\begin{enumerate}
\item A marginal increase from $\beta=0$ lowers both prices and raises consumer surplus at every quality gap if $s\leq1$. If $s>1$, there is exactly one quality gap at which the consumer-surplus response changes from positive to negative.
\item If $s>s_*\simeq1.28657$,\footnote{The threshold is $s_*=\log(2\cosh t_*)$, where $t_*>0$ is the unique solution of $t_*\tanh t_*=1$. It is a sufficient threshold for a both-price reversal.} there is an interval of quality gaps of positive length and a positive interval of price weights starting at zero on which equilibrium is unique and active, both prices rise with $\beta$, and consumer surplus and total surplus fall.
\item For every $s>s_*$, these strict reversals also occur throughout such intervals with sufficiently small positive quality weights and a common increasing, strictly concave advertising technology under which both types advertise. Advertising expenditure and multiplier responses can again be made arbitrarily small. The equilibrium remains unique and active, and every allowed advertising history has a unique price and search continuation.
\end{enumerate}
The comparisons in parts 2 and 3 hold between any two distinct price weights in the respective interval, keeping the other primitives fixed.
\end{proposition}

Consumer losses and higher prices need not occur over the same range. At a fixed search cost above one, consumer surplus falls at all sufficiently large quality gaps, but the low-quality seller's price eventually falls again. Both prices rise over an intermediate range, where the reduction in consumers' selectivity outweighs both direct price cuts. At a reference market with $h_H+h_L=1$, the exact test for this reversal is
\begin{equation}
 (h_L-h_H)\log\frac{h_L}{h_H}>2.
 \label{px:main-rejection-test}
\end{equation}
The logarithm is the difference in inspection gains and the other factor is the difference in rejection probabilities. This is the logistic specialization of~\eqref{eq:px-general-primitive-inequality}. The advertising construction in Appendix~\ref{px:positive-advertising} preserves these strict inequalities by keeping advertising responses small. For other technologies, the response of visits in~\eqref{eq:policy-gain-main} determines whether advertising reinforces or offsets the price reward's reallocation.

An existing quality reward changes the logistic comparison through visits. With equal type masses, no advertising, and $\beta=0$, a fall in consumer surplus requires
\begin{equation}
 s>\exp\{k(v_H-v_L)\}.
 \label{eq:price-quality-protection}
\end{equation}
If this necessary condition fails, consumer surplus rises and both prices fall. At a fixed positive quality weight, sufficiently large quality differences concentrate visits on the better product and restore the consumer benefit. Thus the single-threshold classification in part 1 of Proposition~\ref{px:proposition} is specific to the initially absent quality reward. Appendix~\ref{px:positive-quality-weight} proves these statements.

Price-directed search makes the information in posted prices important for competition \citep{ArmstrongZhou2011,DingZhang2018}, while personalized ranking can raise prices by reducing the return to a price cut \citep{QiuEtAl2025}. The present comparison raises that direct return. Its adverse effect comes from the loss of alternatives that sustain consumers' willingness to reject an unattractive offer. The framework separates the seller's incentive from this market effect. Evaluating a price reward solely by its immediate pricing incentive would omit the products consumers become less likely to encounter.

\subsection{Product guidance and search opportunities}\label{sec:quality}

Price need not identify a better search opportunity. A quality reward instead makes products with better public characteristics easier to encounter. Retain common cost and a common match distribution satisfying Assumption~\ref{ass:match-en}, allow advertising technologies to differ, and raise $k$ at a fixed $\beta\geq0$. The conditional pricing equation~\eqref{px:main-price} does not depend on $k$. Thus the reform works through the distribution of opportunities, the second term in~\eqref{eq:policy-gain-main}.

The common match distribution provides the ordering needed for this comparison. At a given reservation utility, a higher-quality seller raises its conditional price by less than its quality advantage. Its acceptance cutoff therefore falls. Write $S_i=1-F(u+p_i-v_i)$ for its acceptance probability and $G_i=G(u+p_i-v_i)$ for its expected inspection gain. Both $S_i$ and $G_i/S_i$, the expected gain conditional on acceptance, weakly increase with quality. For active search, transaction shares give the useful representation
\begin{equation}
 M(u)=\frac{\E_\omega[G_i/S_i]}{\E_\omega[1/S_i]}.
 \label{eq:quality-transaction-search-gain}
\end{equation}
To see its role, compare two quality weights at the same reservation utility. A seller weakly increases advertising when its quality multiplier rises by more than the common demand scale, and weakly reduces it when the multiplier rises by less. After these choices adjust, the types losing transaction share have lower quality than those gaining it. The numerator in~\eqref{eq:quality-transaction-search-gain} rises weakly and the denominator falls weakly. Search consequently becomes more attractive even with different advertising technologies. The proof extends this argument to advertising entry and exit and to consumer participation boundaries.

\begin{theorem}[A higher quality weight benefits consumers]\label{thm:quality}
Fix the price weight $\beta\geq0$ and increase the nonnegative quality weight. With common cost and a common match distribution with a nondecreasing hazard, the increase weakly raises conditional gains from search. The lowest and highest equilibrium consumer surplus and reservation utility weakly increase. The lowest and highest equilibrium price of each type weakly decrease. These comparisons allow different advertising technologies, zero advertising expenditure, and multiple equilibria.
\end{theorem}

The higher conditional search gain raises the reservation utility needed to restore search indifference. More selective consumers induce lower prices through~\eqref{eq:price-response}. With a unique equilibrium this compares the two market outcomes. With multiple equilibria the theorem orders their extremes, following \citet{MilgromRoberts1994}; Appendix~\ref{app:quality-finite} proves the comparison across all consumer branches. The result combines guidance among heterogeneous firms, as in \citet{Zhang2026}, with the advertising choices studied by \citet{HaanMoraga2011}.

\paragraph{Lower prices and a higher average paid price.}
Lower prices do not settle what happens to the average price paid. That average uses transaction shares, $\E_\omega[p_i]$, which change as consumers find better products. The decomposition in~\eqref{eq:transaction-price-decomposition} separates each seller's price cut from the shift of purchases toward products that remain relatively expensive. The shift can dominate even when every seller cuts its price and consumers gain. A higher average selling price can therefore accompany stronger competition for each product.

For the transaction-price comparison, specialize to two types with the common advertising technology
\begin{equation}
 g(z)=(1+z/\kappa)^\gamma,\qquad 0<\gamma<1,\quad \kappa>0.
 \label{eq:quality-common-advertising}
\end{equation}
The exponent $\gamma$ governs diminishing returns; $\kappa$ sets the advertising expenditure scale. We also use this technology for the match-distribution comparison below. When every firm advertises, the advertising condition~\eqref{eq:advertising-choice-main} implies $z_i=\gamma(p_i-c)D_i-\kappa$, where $D_i$ is the demand for one type-$i$ firm. Aggregating gives
\begin{equation}
 \Ads=\gamma\bigl(\E_\omega[p_i]-c\bigr)-\kappa.
 \label{eq:quality-advertising-accounting}
\end{equation}
Aggregate advertising expenditure therefore rises with the average transaction markup when both types advertise.

For the next comparison, consider two populations, $L$ and $H$, initially
with common quality $v_L$, and then give population $H$ a quality advantage.
Identify the baseline acceptance cutoff $x_s$ by $G(x_s)=s$. It is determined by the common match distribution and search cost.
Primes below denote derivatives with respect to the match cutoff. At equal qualities, the markup is $1/[\beta+h(x_s)]$. The restrictions
$v_L-c+x_s-1/[\beta+h(x_s)]>0$ and
$\kappa<\gamma/[\beta+h(x_s)]$ make search active and advertising strictly
positive in that baseline.

\begin{corollary}[Price cuts with a higher average transaction price]\label{cor:quality-selection}
Consider two types with common cost and a common match distribution, and the
common advertising technology~\eqref{eq:quality-common-advertising}. Fix any
price weight $\beta\geq0$ and any compact interval of nonnegative quality
weights. Suppose the baseline restrictions above hold, the hazard is four
times continuously differentiable near $x_s$, and $h'(x_s)>0$. Let
$\alpha_H$ denote the firm mass of the population receiving a small positive
quality advantage. If
\begin{equation}
 (1-2\alpha_H)\Bigl\{2[h'(x_s)]^2+
 [\beta+h(x_s)]\bigl[h(x_s)h'(x_s)-h''(x_s)\bigr]\Bigr\}>0,
 \label{eq:quality-selection-primitive}
\end{equation}
then, for every sufficiently small positive quality gap, equilibrium is
unique and active and both types advertise throughout the interval.
Increasing $k$ lowers each price, raises consumer surplus, raises the average
transaction price, and increases aggregate advertising.
\end{corollary}

The price cuts follow from the improvement in search opportunities established in Theorem~\ref{thm:quality}. Purchases also move toward the higher-quality type, which remains more expensive. Condition~\eqref{eq:quality-selection-primitive} makes this composition effect larger than the individual price cuts in~\eqref{eq:transaction-price-decomposition}. The average paid price rises because consumers buy better products, although every product becomes cheaper. Equation~\eqref{eq:quality-advertising-accounting} then implies higher advertising expenditure.

A simple sufficient case has $\alpha_H<1/2$ and $h''(x_s)\leq h(x_s)h'(x_s)$. Near equal qualities, shifting visits toward the less numerous high-quality population raises both mean quality and quality dispersion. The distributional curvature determines how this dispersion affects transaction markups. A common increase in all qualities instead raises reservation utility by the same amount and leaves prices unchanged. The comparison thus separates a change in the quality level from a change in purchase composition. The proof gives the local expansion and verifies the condition for logistic and normal matches. If the expression on the left of~\eqref{eq:quality-selection-primitive} is strictly negative, the average transaction price instead falls, while consumer surplus still rises and both individual prices fall.

\subsection{Match distributions and advertising feedback}\label{sec:match-guidance}
With a common match distribution, better public quality weakly improves both the chance of acceptance and the expected gain conditional on acceptance after sellers have priced. Different distributions can separate these advantages. A broadly suitable product may offer a modest improvement when accepted, while another offers a large improvement to the few buyers it suits. Similar expected inspection gains can therefore coexist with different acceptance probabilities. We use this distinction to ask whether rewarding the best inspection still helps consumers when sellers adjust advertising.

Return to the heterogeneous match distributions of Section~\ref{sec:model}, keeping quality $v$ and cost $c$ common. Retain the exponential price component in~\eqref{eq:applications-score}, with fixed price and quality weights. A reform multiplies the score of one publicly identifiable type, $H$, by $e^\eta$, leaving the other types' direct scores unchanged. The target is fixed at the baseline, where it offers the strictly largest inspection gain. Product characteristics and the information learned at inspection remain fixed throughout the comparison.

Write $S_i=S_i(u+p_i-v)$ and $G_i=G_i(u+p_i-v)$ for acceptance and expected inspection gain at the baseline conditional prices. Sellers may differ in how strongly their advertising multipliers respond to a change in operating returns. At a positive choice with $g_i''(z_i)<0$, define the realized response
\begin{equation}
 \chi_i=-\frac{[g_i'(z_i)]^2}{g_i(z_i)g_i''(z_i)}.
 \label{eq:ad-response-definition}
\end{equation}
The operating return per unit of multiplier is the profit before advertising costs that the seller would earn with its multiplier set to one, holding market aggregates fixed. A one-percent increase in this return raises its optimally chosen multiplier by approximately $\chi_i$ percent. Set $\chi_i=0$ when the optimal advertising choice remains zero after a small change, as it does when the initial marginal return is strictly below marginal cost. The response measures how readily a seller expands its exposure, not the size of its advertising budget.

Compare two responses to the same reward from the same baseline. First hold advertising expenditure at its baseline choices, allowing prices and consumer search to adjust. At a fixed reservation utility, conditional prices remain unchanged, and the reward raises the conditional search gain by $\alpha_H\rho_H(G_H-s)>0$ by moving visits toward the best inspection. When advertising also adjusts, the common demand scale changes every seller's return to exposure. Differentiating that scale gives
\begin{equation}
 \frac{\partial M}{\partial\eta}
 =(1+\chi_H)\alpha_H\rho_H
 \left[G_H-s-
 \frac{S_H\Cov_\rho(G_i,\chi_i)}
      {\E_\rho[S_i(1+\chi_i)]}\right].
 \label{eq:match-reform-response}
\end{equation}
The first term is the gain from favoring $H$. The second captures the advertising response to the common demand scale. It is a loss when sellers offering valuable inspections are also responsive advertisers: a higher scale reduces their returns, and their advertising cuts remove valuable opportunities. The scale depends on acceptance probabilities, whereas continued search depends on expected inspection gains. Because these need not rank products in the same way, the loss can exceed the direct gain. Appendix~\ref{app:match-distributions} derives the formula from the policy equations of Section~\ref{sec:core}.

\begin{proposition}[An advertising reversal of product guidance]\label{prop:match-ad-reversal}
Consider an active baseline with $G_H>\max_{i\ne H}G_i$ and the same types choosing positive advertising in a neighborhood of the baseline.
\begin{enumerate}
\item Suppose the conditional search-gain functions with fixed and adjusting advertising both decrease locally in reservation utility. The reform benefits consumers with fixed advertising. With advertising adjusting, it harms consumers exactly when
\begin{equation}
 \frac{S_H\Cov_\rho(G_i,\chi_i)}{\E_\rho[S_i(1+\chi_i)]}>G_H-s.
 \label{eq:match-reversal-test}
\end{equation}
Prices weakly fall in the first comparison and weakly rise in the second; the signs are strict when every cutoff hazard has positive slope.
\item The strict reversal occurs for an open set of three-type primitives with common quality, cost, mean match value, and the common advertising technology~\eqref{eq:quality-common-advertising}. Matches have smooth positive full-support densities and nondecreasing hazards, and the price weight is strictly positive. Both comparisons have unique active equilibrium over a nonempty interval of small reforms. Two types advertise positively, while the third stays at a strict zero-advertising choice.
\end{enumerate}
\end{proposition}

To interpret the condition, let $H$ and another advertising type, $M$, offer the same inspection gain, exceeding that of a nonadvertising type $L$. Under the common power technology, positive advertisers have response $\gamma/(1-\gamma)$, while $L$ has response zero. Advertising reverses the benefit exactly when
\begin{equation}
 \frac{\gamma}{1-\gamma}\,\alpha_M\rho_M(S_H-S_M)
 >\E_\rho[S_i].
 \label{eq:match-conversion-gap}
\end{equation}
The acceptance gap between nearly equally valuable inspections must therefore be large relative to average acceptance, and advertising must respond strongly enough. Making $H$ slightly better preserves the strict inequality and gives the proposition's strictly best target.

The comparison can be generated by match widths, population masses, and the advertising technology. At a zero price weight, take a uniform reference family with widths $a_H=a$, $a_M=1/a$, and $a_L=1$, where $1/3<a<1$, and normalize $v-c-u=1$, with $v-c>1$ for active search. Types $H$ and $M$ then offer equal gains, but $H$ is more often accepted. There is a nonempty interval of advertising scales producing a negative conditional search-gain response if
\begin{equation}
 \frac{\gamma}{1-\gamma}\alpha_M(1-a^2)
 >(1+a)(\alpha_H+a\alpha_M)
   +2a\alpha_L\left[\frac{4a}{(1+a)^2}\right]^{\gamma/(1-\gamma)}.
 \label{eq:match-width-condition}
\end{equation}
Here $a_i$ is the half-width of the zero-mean uniform match distribution. The left side increases with the mass of the displaced advertiser, the acceptance gap, and its advertising response. The right side includes the baseline exposure retained by the nonadvertising type. Search cost is set by the gains in this constructed family. Appendix~\ref{app:match-distributions} derives the advertising-scale interval, makes $H$ strictly better by slightly reducing $a_M$, and verifies decreasing search gains on the entire candidate interval for a subfamily. Small Gaussian smoothing and a positive price weight then give the admissible markets in the proposition.

In these markets, the adjustment begins with advertising rather than prices. At a given reservation utility, the favored advertiser expands its advertising expenditure, while the other valuable advertiser reduces it. The worse, nonadvertising type retains baseline exposure and gains visit share. The loss from this shift can exceed the gain from moving visits between the two valuable types. With no direct conditional price change, the visit term in~\eqref{eq:policy-gain-main} lowers the value of another inspection. Consumers become less selective, and~\eqref{eq:policy-response-core} gives higher equilibrium prices at every seller. Differences in advertising responses are essential: if all types advertise under the common power technology, the covariance in~\eqref{eq:match-reform-response} is zero. With only two types, a pure reward for the better inspection also cannot reverse its consumer benefit on a decreasing branch.

Table~\ref{tab:match-ad-comparison} illustrates the comparison with a positive price weight and smooth full-support matches. The reward raises the target's score by the factor $e^{.1}$. Prices and search adjust in both columns; advertising is either fixed at its baseline choices or reoptimized. The consumer-surplus changes are small, while aggregate advertising falls by about 16 percent when it adjusts. The example also shows why a higher conversion rate---the probability of purchase after an inspection---need not indicate better search opportunities: acceptance rises and search shortens even in the comparison in which consumers lose.

\begin{table}[htbp]
\centering
\caption{Changes under the same recommendation reward}
\label{tab:match-ad-comparison}
\begin{tabular}{lcc}
\toprule
Outcome & Fixed advertising & Adjusting advertising\\
\midrule
Consumer surplus per buyer & $+0.0010$ & $-0.000063$\\
Each price & Falls & Rises\\
Aggregate advertising & Unchanged & $-16.3\%$\\
Acceptance per inspection & $+0.77$ percentage points & $+2.81$ percentage points\\
Expected inspections & $-1.6\%$ & $-5.4\%$\\
\bottomrule
\end{tabular}
\end{table}
With advertising adjusting, the valuable type $M$ loses visit share from $0.6$ to about $0.545$, while the worse type $L$ gains share from $0.2$ to about $0.202$. The latter shift removes enough search value to outweigh the gain from redirecting visits between $M$ and $H$. Appendix~\ref{app:match-illustration} gives the primitives and verifies uniqueness throughout the policy interval.

\citet{Zhou2022} shows how improvements in information or search-pool selection can shorten search, raise prices, and harm consumers. \citet{AndersonRenault2026} connects heterogeneous demands, pricing, and position bids through ordered-search externalities. The present comparison identifies the contribution of advertising adjustment to a given reward. Consumers gain after endogenous repricing with advertising fixed, but lose when advertising adjusts. Match shape separates acceptance probabilities from inspection gains; advertising responses determine which opportunities retain exposure. Equation~\eqref{eq:policy-gain-main} values those opportunities, and~\eqref{eq:policy-response-core} determines prices. Evaluating the targeted product alone would miss the valuable inspections lost through other sellers' responses.

Advertising expenditure must also be distinguished from real resource use. Click payments such as those in Amazon's Sponsored Products are seller costs and platform receipts\footnote{Amazon Ads, ``Sponsored Products''; see footnote~\ref{fn:amazon-ads}.}, whereas managing campaigns can consume real resources. A total-surplus calculation that includes the platform counts these payments as transfers and deducts the real-resource component of advertising.

\section{Discussion and Extensions}\label{sec:extensions}

\subsection{A restriction on optimal recommendation weights}\label{sec:platform-design}

The preceding applications evaluate changes in recommendation. We now let the platform choose the rule. Advertising payments and commissions give it a reason to care about the prices sellers charge and the products consumers buy. The equilibrium construction yields a restriction on how it combines rewards for quality and low prices.

Retain the exponential score~\eqref{eq:applications-score}, two quality types $v_H>v_L$ with common cost $c$, a common match distribution, and the advertising technology~\eqref{eq:quality-common-advertising}. High-quality sellers have mass $\alpha_H$; the other type has mass $1-\alpha_H$. The platform chooses $k,\beta\geq0$ before sellers choose advertising expenditure and then prices. Consumers search last. Advertising expenditure is a payment to the platform, which also receives a fixed fraction $\tau\in(0,1)$ of each sale.\footnote{\interlinepenalty=10000 PDD Holdings Inc., \emph{Annual Report 2025}, p.~104, describes charges for product marketing in search and browsing results, alongside transaction-service fees. \href{https://investor.pddholdings.com/static-files/92dafbdc-3125-4f2c-a28f-3d61203efbaf}{Annual report}. These are consolidated group revenue categories. The fixed ad valorem commission here is a modeling assumption.} Seller profit is $[(1-\tau)p_i-c]D_i-z_i$. Write $\bar p=\E_\omega[p_i]$ for the average transaction price and $R$ for platform cash receipts. The objective is
\begin{equation}
 J_\theta(k,\beta)=R(k,\beta)+\theta\CS(k,\beta),\qquad
 R=\Ads+\tau\bar p,\qquad \theta\geq0.
 \label{eq:platform-objective}
\end{equation}
The weight $\theta$ measures the platform's value of consumer surplus in addition to current receipts, for example through customer retention. The consumer population is initially fixed.

The seller problem has the form studied in Section~\ref{sec:core}, with effective cost $c/(1-\tau)$ and advertising units $z/(1-\tau)$. Conditional markups are $p_i-c/(1-\tau)=1/[\beta+h(u+p_i-v_i)]$; advertising and search then determine the market outcome. When both types advertise, their first-order conditions imply
\begin{equation}
 R=\bigl[\tau+\gamma(1-\tau)\bigr]\bar p-\gamma c-\kappa.
 \label{eq:platform-revenue-interior}
\end{equation}
Thus both revenue sources increase with the average transaction price when both types advertise. The transaction decomposition in Section~\ref{sec:core} identifies two ways the platform can raise that price: sellers can charge more, or purchases can shift toward more expensive products. Search consistency determines which combinations preserve consumer surplus. The result below also allows one or both types to choose zero advertising.

Maintain active search and compare the equilibrium outcomes that feasible recommendation weights can implement. To interpret this comparison as a choice between rules, assume a locally unique market response as the weights change.

\begin{proposition}[A restriction on optimal recommendation weights]\label{prop:design-positive-beta}
Under the preceding two-type specification, suppose the common match distribution has a differentiable, nondecreasing hazard. Every active outcome induced by finite $k>0$ and $\beta>0$ admits a nearby feasible equilibrium with the same consumer surplus and strictly higher cash receipts. Thus, for every $\theta\geq0$, any finite optimum over implementable outcomes satisfies $k^*\beta^*=0$. The same conclusion holds for optimal rules with locally unique continuation, including positive endogenous consumer participation when consumers differ only in their outside options.
\end{proposition}

The argument starts by reducing the price weight at a given reservation utility. The conditional pricing equation then makes every seller raise its price. To leave consumers equally willing to search, the platform must change the mix of products they encounter. With a nondecreasing hazard, the required adjustment increases the higher-quality type's share of purchases. That type charges a weakly higher price, so the shift in purchases reinforces the rise in individual prices. Commission receipts strictly increase.

Advertising receipts weakly increase along the same adjustment. Under the common technology, high-quality sellers advertise at least as much as low-quality sellers. If both types advertise, receipts rise with the average transaction markup; if only the high-quality type advertises, its larger transaction share and markup raise receipts. These comparisons agree when a type enters or leaves advertising. Starting with a positive quality weight allows a small adjustment of that weight to implement the new composition. Consumers obtain the same surplus and the platform earns more, regardless of $\theta$. Appendix~\ref{app:platform-design} proves the construction. Consequently, a finite optimum with a positive price reward must place the quality weight at its lower bound.

If consumers differ only in their outside options, preserving reservation utility also preserves entry and each entrant's surplus. The same improvement therefore applies with endogenous consumer participation. Seller participation introduces a different constraint: the shift of purchases toward high quality can reduce low-quality sellers' profits. Strictly slack seller participation constraints allow the local adjustment; binding constraints require an entry analysis.

\subsection{When advertising changes the reward to a price cut}

Separability makes advertising multiply the return at every price by the same factor. For a joint score, advertising can instead change the exposure gained from a price cut. Let $w_i(p,z)$ be positive and twice continuously differentiable, nonincreasing and log-concave in price, and increasing in advertising. With a differentiable match hazard, the cross derivative $\partial^2\log w_i/\partial p\partial z$ determines the resulting conditional price response.

The conditional optimal price $p_i(z,u)$ is still unique at each advertising choice. Its response to advertising is
\begin{equation}
 \frac{\partial p_i}{\partial z}
 =\frac{\partial^2\log w_i/\partial p\partial z}
 {(p_i-c_i)^{-2}+h_i'(u+p_i-v_i)-\partial^2\log w_i/\partial p^2}.
 \label{eq:nonseparable-price}
\end{equation}
All score derivatives are evaluated at $(p_i(z,u),z)$. The denominator is positive. A negative cross derivative means that additional advertising makes a price increase lose a larger proportion of visits; the seller responds by cutting its price. A positive cross derivative weakens that exposure loss and raises the conditional price.

The price adjustment also changes advertising incentives. Reoptimizing price adds a nonnegative term to the curvature of operating returns in advertising. Concavity of the score in advertising at a fixed price is therefore insufficient to ensure diminishing returns after repricing. Appendix~\ref{app:nonseparable} derives this term and~\eqref{eq:nonseparable-price}. This calculation identifies the additional restriction needed when applying the conditional-choice approach to a joint score.

\subsection{A large-assortment demand foundation}

When consumers inspect only a small part of a large assortment, removing products already inspected changes the remaining opportunities little. Consider a finite market with $N$ sellers and a mass $N$ of consumers. After each inspection, the platform draws the next seller from the unvisited set in proportion to its score. Consumers observe independent match values, remember previous offers, and choose optimally whether to inspect another seller, buy an inspected product, or exit.

Fix a public profile of prices and scores. Assume that scores are bounded above and away from zero and match densities are bounded, uniformly as the number of sellers grows. Also require that the expected gain from a first paid inspection exceed its cost by a uniform positive margin, and that there be a common utility level above which every seller's expected improvement is at most half the search cost. The score bounds make each seller's share of visits small; the entry and tail restrictions keep search active while bounding its duration. Appendix~\ref{app:finite} states these uniform bounds explicitly.

Let $\widehat u_N$ be the reservation utility in the stationary comparison market, which draws from the full set after each inspection. For seller $j$, let $d_j^{\mathrm{opt}}$ be its purchase probability under optimal finite search. Under the preceding conditions,
\begin{equation}
 N d_j^{\mathrm{opt}}
 =\frac{w_j[1-F_j(\widehat u_N+p_j-v_j)]}
 {N^{-1}\sum_{\ell=1}^N w_\ell[1-F_\ell(\widehat u_N+p_\ell-v_\ell)]}
 +o(1).
 \label{eq:finite-main}
\end{equation}
The factor $N$ converts a purchase probability into demand from the mass of consumers, and $o(1)$ denotes an error tending to zero as $N$ grows. The approximation is uniform across sellers and unilateral price deviations satisfying the bounds. It therefore approximates the individual demand relevant for pricing, as well as demand at the initial profile.

Search cost bounds the expected number of remaining inspections after any history. Removing the few products already inspected changes continuation opportunities only slightly, and optimal stopping decisions remain close to those in the stationary market. Appendix~\ref{app:finite} uses these observations to establish~\eqref{eq:finite-main}. The result provides a demand foundation for the continuum analysis at public action profiles; strategic equilibrium is characterized in Section~\ref{sec:core}.

\section{Conclusion}

A recommendation rule affects both the seller receiving a visit and the alternatives available to consumers who continue searching. The framework developed here brings these effects into a common equilibrium construction. Conditional pricing and advertising determine the products consumers encounter; search consistency determines how selective consumers are and closes the market. The same construction accommodates differences in quality, costs, matches, and advertising technologies.

The three applications separate incentives to obtain visits from competition through continued search. Stronger price rewards can remove valuable alternatives and raise every price. Quality guidance can lower prices at all sellers while shifting purchases toward more expensive products. With different match distributions, advertising adjustment can overturn the benefit of rewarding the best inspection. These comparisons also distinguish changes in individual prices from changes in purchase composition, and the probability of making a purchase from the value of the alternatives still available.

Recommendation design provides a further use of the framework. In the two-type extension, a feasible change from jointly positive weights preserves consumer surplus and raises platform receipts. This gives a restriction on finite optima over implementable outcomes, using the same price and purchase-selection responses as the applications. Endogenous seller entry and platform competition would add participation and market-share responses to these choices. Studying how platforms jointly choose fees and recommendation under those conditions is a natural next step.

\label{end:main}

\clearpage
\begingroup
\raggedright

\endgroup

\clearpage
\appendix

\setcounter{equation}{0}
\setcounter{figure}{0}
\setcounter{table}{0}
\setcounter{theorem}{0}
\setcounter{proposition}{0}
\setcounter{lemma}{0}
\setcounter{corollary}{0}
\setcounter{assumption}{0}
\setcounter{remark}{0}
\renewcommand{\theequation}{A.\arabic{equation}}
\renewcommand{\thefigure}{A.\arabic{figure}}
\renewcommand{\thetable}{A.\arabic{table}}
\renewcommand{\thetheorem}{A.\arabic{theorem}}
\renewcommand{\theproposition}{A.\arabic{proposition}}
\renewcommand{\thelemma}{A.\arabic{lemma}}
\renewcommand{\thecorollary}{A.\arabic{corollary}}
\renewcommand{\theassumption}{A.\arabic{assumption}}
\renewcommand{\theremark}{A.\arabic{remark}}
\def\theHequation{appendix.\arabic{equation}}
\def\theHfigure{appendix.\arabic{figure}}
\def\theHtable{appendix.\arabic{table}}
\def\theHtheorem{appendix.\arabic{theorem}}
\def\theHproposition{appendix.\arabic{proposition}}
\def\theHlemma{appendix.\arabic{lemma}}
\def\theHcorollary{appendix.\arabic{corollary}}
\def\theHassumption{appendix.\arabic{assumption}}
\def\theHremark{appendix.\arabic{remark}}

\section*{Appendices}
The proofs follow the order of the main analysis. Appendices~\ref{app:basic}--\ref{app:policy-proof} cover demand, equilibrium, and comparative statics; Appendices~\ref{px:appendix}--\ref{app:quality} cover the three recommendation applications and verify Table~\ref{tab:match-ad-comparison}. Appendix~\ref{app:platform-design} proves the design result, and Appendices~\ref{app:nonseparable}--\ref{app:finite} derive the joint-score and finite-assortment results.

\section{Proofs of Lemma~\ref{lem:conditional-price} and Theorem~\ref{thm:aggregation-en}}
\label{app:basic}\label{app:sequential}\label{app:sequential-implementation}

We first prove the conditional pricing result. We then derive demand in every consumer-participation branch, solve the advertising and scale equations, and establish existence and sequential implementation. Throughout, the score is separable and $g_i(0)=1$.

In Appendices~\ref{app:basic}--\ref{app:policy-proof}, derivative subscripts denote partial derivatives of conditional quantities, with the recommendation rule fixed when varying $u$. Write $S_i(x)=1-F_i(x)$ and $Z=\sum_i\alpha_iw_i$; $S_i,h_i,G_i$ without displayed arguments are evaluated at the current acceptance cutoff. Define the optimized price-profit numerator by
\[
 V_i(u;\eta)=(p_i-c_i)w_i(p_i,0;\eta)S_i(u+p_i-v_i).
\]
Operating profit multiplies this numerator by the advertising score multiplier and divides by the common demand scale. The function $\zeta_i$ describing optimal advertising is defined below. Write the price penalty introduced in the uniqueness test as
\[
 b_i(p;\eta)=-\partial_p\log w_i(p,0;\eta).
\]
Its price derivative is $b_{ip}$. When the policy index is suppressed, $b_i=b_i(p_i)$.

\subsection{Proof of Lemma~\ref{lem:conditional-price}}
\begin{proof}
The price objective is $(p-c_i)w_i(p,0)S_i(u+p-v_i)$. Its log derivative is
\[
 (p-c_i)^{-1}-b_i(p)-h_i(u+p-v_i).
\]
It strictly decreases from positive infinity and is negative for large $p$: the positive nondecreasing hazard is bounded below on every right half-line. Integrating that bound gives an exponential survival-tail bound, so the objective tends to zero. Its unique critical point is its unique global optimum $p_i(u)$. This proves the uniqueness and independence assertions of Lemma~\ref{lem:conditional-price}.

Under the stated comparative-static smoothness, implicit differentiation and the envelope theorem give
\[
 \partial_u p_i=-\frac{h_i'(x_i)}{(p_i-c_i)^{-2}+h_i'(x_i)+b_{ip}(p_i)}\in(-1,0],
 \qquad \partial_u x_i=1+\partial_u p_i>0.
\]
The optimized numerator satisfies
\[
 \partial_u\log V_i=-h_i,\qquad V_i=(p_i-c_i)w_i(p_i,0)S_i.
\]
Continuity of $p_i$ follows from the strictly decreasing pricing residual, without the additional smoothness.

\end{proof}

\subsection{Proof of Theorem~\ref{thm:aggregation-en}}
\paragraph{Consumer search and demand.}

For fixed prices and visit shares, let $X$ be utility from a fresh inspection. Write $(x)_+=\max\{x,0\}$ for the positive part of $x$, and define $\mathcal G(y)=\E(X-y)_+=\E_{\rho}G_i(y+p_i-v_i)$. This function is continuous, strictly decreasing, and tends to zero. If $\mathcal G(0)>s$, its unique positive root $\mathcal G(u)=s$ gives the paid-search value $v(y)=\max\{y,u\}$ for a current best $y\ge0$. Indeed,
\[
-s+\E v(\max\{y,X\})=
\begin{cases}u,&y<u,\\y+\mathcal G(y)-s\le y,&y\ge u.\end{cases}
\]
The policy stops on the first offer reaching $u$. Its duration is geometric with positive success probability, so it attains this Bellman value. If $\mathcal G(0)<s$, immediate stopping is optimal after the free inspection. At equality, any stationary probability $\vartheta\in[0,1]$ of continuing after failure is optimal; set $\vartheta=1$ when $u>0$.

With $x_i=u+p_i-v_i$, define $d_i=1-\vartheta+\vartheta S_i(x_i)$ as the probability that search terminates after inspecting type $i$, through acceptance or exit. Its visit-weighted average is $q=\E_{\rho}d_i>0$, the termination probability per draw. Summing visits over search rounds yields
\begin{equation}
 D_i=\frac{\rho_iS_i(x_i)}q=\frac{w_iS_i(x_i)}C,
 \qquad C=Zq=\sum_i\alpha_iw_id_i.
 \label{eq:demand-boundary}
\end{equation}
These demand formulas also give the surplus accounting in every participation branch. The probability of a purchase is $\E_{\rho}S_i/q$, while expected inspections equal $1/q$. A paid inspection occurs only after an unsuccessful draw and a decision to continue, so the expected number of paid inspections is $\vartheta(1-\E_{\rho}S_i)/q$. The expected utility of the accepted offer, before search costs and including the probability of no purchase, is $[\E_{\rho}G_i+u\E_{\rho}S_i]/q$. Therefore
\begin{equation}
 \CS=\frac{\E_{\rho}G_i+u\E_{\rho}S_i-s\vartheta(1-\E_{\rho}S_i)}q,
 \qquad
 \Pi=\frac{\E_{\rho}[(p_i-c_i)S_i]}q-\sum_i\alpha_i z_i,
 \qquad W=\CS+\Pi.
 \label{eq:boundary-accounting}
\end{equation}
If $u>0$, then $\vartheta=1$, $q=\E_{\rho}S_i$, and $\E_{\rho}G_i=s$, yielding $\CS=u+s$ and unit total purchases. At boundary indifference, $u=0$ and $\E_{\rho}G_i=s$, and the numerator of consumer surplus equals $sq$, giving $\CS=s$ for every optimal continuation probability. With only the free inspection, $\vartheta=0$ and $q=1$, giving $\CS=\mathcal G(0)$. Under active search the normalized transaction shares are $\omega_i=\alpha_iw_iS_i/C$; substituting them in the profit expression gives~\eqref{eq:welfare}. At the participation boundary, the unnormalized weights $\alpha_iw_iS_i/C$ instead sum to the purchase probability, which is why the general formula is written before imposing unit purchases.

\paragraph{Advertising and the common scale.}
For a positive operating-return coefficient $r$, the objective $r g_i(z)-z$ is strictly concave and eventually has negative derivative. Its unique maximizer $\zeta_i(r)$ is zero if $rg_i'(0+)\le1$ and otherwise solves $rg_i'(z)=1$. Comparing maximization inequalities at two values of $r$ proves monotonicity; the first-order characterization, including its zero threshold, proves continuity. Consequently the full conditional scale equation
\begin{equation}
 C=\sum_i\alpha_iw_i(p_i,0)d_i
       g_i\!\left(\zeta_i(V_i/C)\right)
 \label{eq:scale-boundary}
\end{equation}
has exactly one positive solution: its right side is continuous and nonincreasing, is bounded below by the positive zero-advertising contribution, and approaches that finite contribution as $C\to\infty$. The left side increases from zero to infinity.

For continuity, choose two positive values of $C$ on opposite sides of the root. The residual has strict opposite signs at these endpoints. Continuity of prices and advertising expenditure choices preserves those signs after a small change in the consumer state; monotonicity keeps the new root inside the same bracket. Taking arbitrarily small brackets proves continuity of the root. Advertising expenditure choices, scores, visit shares, and search gains are then continuous by composition. This argument also covers a firm starting to advertise, because $\zeta_i$ is continuous at its zero threshold. At $\vartheta=1$, the equation reduces to~\eqref{eq:scale-main}.

\paragraph{Existence and sequential implementation.}

At the conditional choices, extend the active-search gain function to $M(u,\vartheta):=\E_{\rho}G_i(x_i)$. The continuation probability $\vartheta$ affects the boundary scale and hence the visit shares. The case $M(u,1)$ is the active-search function $M(u)$ in~\eqref{eq:conditional-search-gain}. Consumer optimality gives exactly
\begin{align}
 u>0:&\quad\vartheta=1,\quad M(u,1)=s;\notag\\
 u=0:&\quad\vartheta=0,\quad M(0,0)\le s;\label{eq:equilibrium-branches}\\
 u=0:&\quad0<\vartheta\le1,\quad M(0,\vartheta)=s.\notag
\end{align}
Since $p_i\ge c_i$, $M(u,\vartheta)\le\max_iG_i(u+c_i-v_i)\to0$, uniformly in shares and $\vartheta$. If $M(0,0)\le s$, the second branch applies. Otherwise, $M(0,1)>s$ gives a positive root by continuity, equality gives $(0,1)$, and $M(0,1)<s$ gives a boundary root in $(0,1)$. Thus at least one consistent outcome exists.

We now implement every such outcome with advertising expenditure chosen before prices. The construction gives a price and search continuation after every advertising history, verifies that the action bounds do not constrain a best reply, and then checks individual advertising deviations.

Let $\bar p_i$ and $\bar z_i^{\rm act}$ be the upper bounds on the feasible price and advertising expenditure of type $i$. Choose these action ceilings to satisfy
\[
 \bar p_i>c_i+\frac1{h_i(c_i-v_i)},\qquad
 g_i'(\bar z_i^{\rm act})<\alpha_i g_i(0)h_i(c_i-v_i).
\]
The price ceiling never binds: $u\ge0$ implies $x_i\ge c_i-v_i$ and $(p_i-c_i)\le1/h_i(c_i-v_i)$.

For any measurable advertising expenditure history, let $\nu_i$ be its within-type advertising expenditure distribution and $\bar g_i=\int g_i(z)\,d\nu_i(z)>0$. Thus $\nu_i$ assigns to a set of advertising expenditures the fraction of type-$i$ sellers choosing values in that set. Bounded feasible advertising expenditure ensures that this integral is finite. Advertising expenditure is sunk in the price subgame and enters the price objective multiplicatively. Conditional prices therefore remain $p_i(u)$ for every firm of a type, even when their advertising expenditures differ. The resulting search gains are
\[
 M_{\boldsymbol\nu}(u)=
 \frac{\sum_i\alpha_iw_i(p_i(u),0)\bar g_i
                  G_i(u+p_i(u)-v_i)}
      {\sum_i\alpha_iw_i(p_i(u),0)\bar g_i}.
\]
The denominator is positive, and the same conditional-pricing continuity proves continuity in $u$. Its weighted-average form gives $M_{\boldsymbol\nu}(u)\le\max_iG_i(u+c_i-v_i)$, independently of the advertising expenditure distribution. If its value at zero exceeds $s$, select a positive root; otherwise select zero and the appropriate consumer continuation. Individual price optimization and optimal search therefore give a continuation equilibrium after every advertising expenditure history.

Such selections can depend measurably only on $(\nu_i)_i$. On the product of probability measures over the compact advertising expenditure intervals, the gain function is jointly continuous. Reservation roots, including zero when its gain is at most $s$, form a nonempty compact correspondence with closed graph; they lie below $U=\max\{0,\max_i[(v_i-c_i)+G_i^{-1}(s/2)]\}$. Its largest-element selection is measurable. Changing that selection at a specified aggregate history preserves measurability and permits any target root and boundary tie rule.

At each price continuation, $d_i\ge S_i$ and $\bar g_i\ge g_i(0)$ imply
\[
 C\ge\alpha_iw_i(p_i,0)g_i(0)S_i,\qquad
 \frac{V_i}C\le\frac1{\alpha_i g_i(0)h_i(c_i-v_i)}.
\]
The chosen advertising expenditure ceiling therefore never binds a best reply. In particular, at or above that ceiling,
\[
 \frac{V_i}C g_i'(z)
 \le\frac{g_i'(z)}{\alpha_i g_i(0)h_i(c_i-v_i)}<1,
\]
so additional advertising expenditure strictly reduces profit. This bound is needed only at a price-equilibrium continuation; advertising expenditure need not have been optimal at the preceding history. At arbitrary realized price profiles, compact actions and positive scores still give a well-defined consumer problem.

An infinitesimal seller's deviation changes neither the aggregate history nor its selected continuation. After deviating in advertising expenditure it reoptimizes its price, obtaining $V_i g_i(z)/C-z$. Its unique best advertising expenditure is precisely $\zeta_i(V_i/C)$. Select the target continuation at a consistent outcome and the constructed continuations elsewhere; consumers search optimally after every realized price profile. This gives a subgame-perfect equilibrium. Conversely, aggregate-invariant continuation makes these same conditional best replies necessary in any pure equilibrium. Their uniqueness forces identical actions within type and the scale and branch equations above. This proves Theorem~\ref{thm:aggregation-en} in the stated continuation class.

\section{Proof of Proposition~\ref{thm:uniqueness}}
\label{app:uniqueness}

We verify the candidate interval, establish regularity across advertising entry and exit, and bound the reallocation term in~\eqref{eq:feedback}. Suppress the policy index. For each type, let $x_i^s$ be the required match giving expected improvement $s$, let $p_i^s$ be its optimal price at that cutoff, and let $u_i^s$ be the associated reservation utility:
\begin{equation}
 x_i^s=G_i^{-1}(s),\qquad
 \frac1{p_i^s-c_i}=h_i(x_i^s)+b_i(p_i^s),\qquad
 u_i^s=v_i-p_i^s+x_i^s.
 \label{eq:type-search-levels}
\end{equation}
The superscript $s$ denotes indifference at search cost $s$. These individual calculations recover the candidate interval $I$ defined in the main text by~\eqref{eq:root-interval}.
By Lemma~\ref{lem:conditional-price}, each conditional type gain strictly decreases with $u$ and equals $s$ at $u_i^s$. All gains therefore exceed $s$ below the lower endpoint of $I$ and are below $s$ above its upper endpoint. Every active equilibrium lies in $I$. Strict entry, $\min_i u_i^s>0$, excludes the participation boundary because every type's gain at zero exceeds $s$. The same statements apply after fixed advertising histories, which change weights but not conditional prices.

The quantities $\Delta_G$, $\Delta_J$, and $\bar\chi$ are defined in the main text. On compact $I$, the strict inequality~\eqref{eq:primitive-unique} is equivalent to the existence of a constant $\kappa\in[0,1)$ such that
\[
 \Delta_G(u)\Delta_J(u)
 \leq4\kappa\min_i\{S_i(x_i(u))[1+\partial_u p_i(u)]\}
 \quad\text{for every }u\in I.
\]
Indeed, both sides of the strict test are continuous and its right side is strictly positive. We use this uniform margin below.

\paragraph{Advertising bounds and regularity.}

The bounds used in the main text can be written using the advertising choice function $\zeta_i$ as
\begin{align}
 C^{\rm org}(u)&=\sum_j\alpha_jw_j(p_j,0)S_jg_j(0),\notag\\
 \bar z_i&=\zeta_i\!\left(\max_{u\in I}\frac{V_i(u)}{C^{\rm org}(u)}\right),
 &\bar\chi_i&=\sup_{0<z\le\bar z_i}\frac{-g_i'(z)^2}{g_i(z)g_i''(z)}.
 \label{eq:response-bound-main}
\end{align}
Set $\bar\chi_i=0$ when $\bar z_i=0$. By assumption, $g_i\in C^2$, $g_i''<0$ on each nonempty displayed interval, and $\bar\chi=\max_i\bar\chi_i<\infty$. Since $C\ge C^{\rm org}$, every conditional advertising expenditure is at most $\bar z_i$. The assumed price and hazard smoothness applies on the compact conditional domains generated by $I$.

At positive advertising expenditure write
\begin{equation}
 \lambda_i=-g_i'/g_i'',\qquad
 \chi_i=-g_i'^2/(g_i g_i'');
 \label{eq:advertising-response}
\end{equation}
set effective responses to zero on fixed zero-advertising branches. Conditional scales and scores are Lipschitz on $I$, including entry points. This regularity allows us to integrate the derivative bound below across changes in advertising participation. To establish it, cap advertising expenditure at $\bar z_i$ and define the capped log-multiplier function $E_i(r)=\log g_i(\min\{\zeta_i(e^r),\bar z_i\})$, where $r$ is the log operating-return coefficient in this argument. On its interior branch $E_i'=\chi_i$; on capped or zero branches its derivative is zero. Continuity and the finite bound make $E_i$ globally Lipschitz. The capped and original scale equations have the same unique root. The residual written in log-scale coordinates is
\[
 \mathcal F(\log C,u)=\log C-\log\sum_i\alpha_iw_i(p_i,0)S_i
          \exp E_i(\log V_i-\log C).
\]
It is locally Lipschitz and $\mathcal F(\log C',u)-\mathcal F(\log C,u)\ge \log C'-\log C$ for $C'>C$. For example, if $\log C(u')\ge \log C(u)$, monotonicity and the root identities give
\begin{align*}
 \log C(u')-\log C(u)
 &\le \mathcal F(\log C(u'),u)-\mathcal F(\log C(u),u)\\
 &=\mathcal F(\log C(u'),u)-\mathcal F(\log C(u'),u')\\
 &\le K_{\rm Lip}|u'-u|.
\end{align*}
Here $K_{\rm Lip}$ is a finite local Lipschitz bound, independent of $u,u'$ within the neighborhood. The reverse ordering is treated identically. The zero-advertising lower bound and the finite upper bound $C\le\sum_i\alpha_iw_i(p_i,0)S_i g_i(\bar z_i)$ keep the solution graph in a compact region. A finite neighborhood cover therefore makes the Lipschitz constant uniform on $I$. Positive scores and normalization transfer this property to shares and $M$.

This argument concerns advertising score multipliers and the common scale. On the open set where advertising expenditure is positive, the smooth first-order condition permits differentiation wherever the common scale is differentiable. On the zero-advertising set, the log-multiplier function equals the constant $\log g_i(0)$, so its derivative is zero almost everywhere on that set. Intersecting these full-measure differentiability sets across the finitely many types gives the response formulas below almost everywhere on $I$, with $\chi_i=0$ on zero-advertising sets. Absolute continuity therefore permits integration across entry and exit points, without requiring their number to be finite or the active advertising response to have a limit at zero.

\paragraph{The reallocation bound.}

Differentiating $g_i'(z_i)=C/V_i$ and using $(\log V_i)_u=-h_i$ gives $\partial_u z_i=-\lambda_i(h_i+\partial_u\log C)$ at a positive choice. Since $(\log w_i(p_i,0))_u=b_i(-\partial_u p_i)$, the log-score response is $-b_i\partial_u p_i-\chi_i(h_i+\partial_u\log C)$. Transaction shares satisfy $\omega_i=\alpha_iw_iS_i/C$, so
\begin{align*}
 \partial_u\log\omega_i
 &=-b_i\partial_u p_i-\chi_i(h_i+\partial_u\log C)
                   -h_i(1+\partial_u p_i)-\partial_u\log C\\
 &=-(1+\chi_i)(h_i+\partial_u\log C)
                       -\frac{\partial_u p_i}{p_i-c_i}.
\end{align*}
Their sum is one; hence the transaction-share average of this derivative is zero. Rearranging gives
\begin{align}
 -\partial_u\log C
 &=\frac{\E_\omega[(1+\chi_i)h_i+\partial_u p_i/(p_i-c_i)]}
             {\E_\omega(1+\chi_i)},\notag\\
 \partial_u\log w_i&=-b_i\partial_u p_i-\chi_i(h_i+\partial_u\log C).
 \label{eq:H-main}
\end{align}
We can now bound the response of the common scale without solving for advertising choices. The price identity $(p_i-c_i)^{-1}=h_i+b_i$ shows that $-\partial_u\log C$ is a weighted average of $h_i$ and $h_i(1+\partial_u p_i)-b_i(-\partial_u p_i)$. Thus, with $d_-=\min_i[h_i(1+\partial_u p_i)-b_i(-\partial_u p_i)]$ and $h_+=\max_i h_i$, we have $d_-\le (-\partial_u\log C)\le h_+$. Define the ranges $\Delta_G=\max_iG_i-\min_iG_i$ and $\Delta_b=\max_i b_i(-\partial_u p_i)-\min_i b_i(-\partial_u p_i)$. The bound~\eqref{eq:visibility-response-bound} is $\Delta_J=\Delta_b+\bar\chi(h_+-d_-)$ and bounds the range of log-score responses. Indeed, if $(-\partial_u\log C)\le\min_i h_i$, the advertising terms lie in $[-\bar\chi(h_+-(-\partial_u\log C)),0]$; otherwise they lie in $[-\bar\chi(h_+-(-\partial_u\log C)),\bar\chi((-\partial_u\log C)-\min_i h_i)]$. Either interval has length at most $\bar\chi(h_+-d_-)$.

The score-response bound controls how changes in the consumer state reallocate visits. For visit shares, normalization gives
\[
 \alpha_i\partial_u\rho_i=\alpha_i\rho_i\left[\partial_u\log w_i-\E_{\rho}(\partial_u\log w_i)\right].
\]
Combining this expression with $(G_i(x_i))_u=-S_i(1+\partial_u p_i)$ proves~\eqref{eq:feedback}. For any random variable in an interval of length $d$, its variance is at most $d^2/4$. Cauchy--Schwarz therefore bounds the reallocation covariance by $\Delta_G\Delta_J/4$, regardless of the inspection composition. Consequently
\[
 M_u\le-\E_{\rho}[S_i(1+\partial_u p_i)]+\frac14\Delta_G\Delta_J
 \le-(1-\kappa)\min_i S_i(1+\partial_u p_i)<0
\]
under~\eqref{eq:primitive-unique}. The final minimum is uniformly positive on compact $I$. Absolute continuity therefore makes $M$ strictly decreasing across transitions into and out of positive advertising.

At the endpoints of $I$ all type gains are respectively at least and at most $s$. Existence and uniqueness follow; a singleton interval directly determines the root. Strict entry excludes boundary outcomes, and conditional uniqueness determines all remaining choices.

For every fixed advertising expenditure history the score derivative is simply $-b_i\partial_u p_i$. Dropping the nonnegative advertising term from~\eqref{eq:visibility-response-bound} gives the same strict-decrease argument for its pricing continuation. The branch bracket and strict entry still apply. Thus these continuations are unique as well. Together with Appendix~\ref{app:basic}, this proves Proposition~\ref{thm:uniqueness}.

\paragraph{Verification of the two sufficient cases.}\label{cor:au-smallchi}
For the small-response case stated after Proposition~\ref{thm:uniqueness}, define the remaining margin without advertising responses and its coefficient by
\[
 \eta_0=\inf_{u\in I}\{4\min_iS_i(1+\partial_u p_i)-\Delta_G\Delta_b\}>0,
 \qquad K_0=\sup_{u\in I}\Delta_G(h_+-d_-)<\infty.
\]
Then $\bar\chi K_0<\eta_0$ implies~\eqref{eq:primitive-unique}; if $K_0=0$, no further bound is needed. With price-independent scores, $b_i=0$ and $\eta_0=4\min_{u,i}S_i(1+\partial_u p_i)>0$. Small responses allow positive advertising: for positive technology constants $\delta_i,\epsilon_i,\Lambda_i$, the function $g_i(z)=1+(\epsilon_i/\delta_i)\log(1+\Lambda_i z)$ has $\chi_i\le\epsilon_i/\delta_i$, while $\Lambda_i$ controls its initial marginal return independently.

\label{cor:au-nearhomogeneous}
For the nearby-opportunity case, take strict-entry perturbations whose intervals lie in a common compact set, whose $S_i(1+\partial_u p_i)$ stay uniformly positive, and whose price, hazard, and advertising responses are uniformly bounded. If $\Delta_G$ tends uniformly to zero, the left side of~\eqref{eq:primitive-unique} tends to zero and its right side stays positive. The uniqueness conclusion follows for all sufficiently small perturbations.

\section{Proof of Lemma~\ref{thm:policy} and the response decompositions}
\label{app:policy-proof}

Work on a differentiable active-search branch on which the set of types choosing positive advertising expenditure is unchanged. A partial policy derivative holds $u$ fixed after conditional firm choices adjust. A total derivative also includes the equilibrium change in $u$. Search cost is fixed throughout.

The repeated weight calculation can be stated once. Let $r_i(y)>0$ be differentiable unnormalized weights, $\pi_i=\alpha_i r_i/\sum_j\alpha_j r_j$ their normalized counterparts, and $Y_i(y)$ a differentiable type-specific quantity. Then
\begin{align}
 \partial_y\pi_i&=\pi_i\bigl[\partial_y\log r_i-\E_\pi(\partial_y\log r_i)\bigr],\notag\\
 \partial_y\E_\pi Y_i&=\E_\pi(\partial_y Y_i)
       +\Cov_\pi(Y_i,\partial_y\log r_i).
 \label{eq:normalized-expectation}
\end{align}
The subtracted average keeps the weights summing to one. Visits use $r_i=w_i$, whereas transactions use $r_i=w_iS_i$.

To distinguish a direct score change from the response after repricing, define only within this appendix
\[
 \phi_i=\left.\partial_\eta\log w_i(p,0;\eta)\right|_{p=p_i}.
\]
This derivative holds the firm's price fixed. The price envelope gives $\partial_\eta\log V_i=\phi_i$ and $\partial_u\log V_i=-h_i$. With the advertising expenditure response $\lambda_i=-g_i'/g_i''$ from Appendix~\ref{app:uniqueness}, the advertising first-order condition yields
\[
 \partial_\eta z_i=\lambda_i(\phi_i-\partial_\eta\log C),\qquad
 \partial_u z_i=-\lambda_i(h_i+\partial_u\log C).
\]
Differentiating transaction shares and using the price condition gives
\[
 \partial_\eta\log\omega_i
 =(1+\chi_i)(\phi_i-\partial_\eta\log C)
       -\frac{\partial_\eta p_i}{p_i-c_i}.
\]
Transaction shares must continue to sum to one, so their transaction-weighted average log derivative is zero. This condition determines the scale response. Differentiating scores and search gains then gives
\begin{align}
 \partial_\eta\log C
 &=\frac{\E_\omega[(1+\chi_i)\phi_i]
             -\E_\omega[\partial_\eta p_i/(p_i-c_i)]}
            {\E_\omega(1+\chi_i)},\notag\\
 \partial_\eta\log w_i
 &=\phi_i-b_i\partial_\eta p_i
                   +\chi_i(\phi_i-\partial_\eta\log C),
 \label{eq:policy-scale}\\
 \partial_\eta M
 &=-\E_{\rho}(S_i\partial_\eta p_i)
       +\Cov_{\rho}(G_i,\partial_\eta\log w_i).
 \label{eq:policy-N}
\end{align}
Implicit differentiation of the search equation yields
\begin{align}
 \frac{\dd u}{\dd\eta}
 &=-\frac{\partial_\eta M}{\partial_u M},\notag\\
 \frac{\dd p_i}{\dd\eta}
 &=\partial_\eta p_i+\partial_u p_i\frac{\dd u}{\dd\eta},\qquad
 \frac{\dd z_i}{\dd\eta}
 =\lambda_i\frac{\dd\log(V_i/C)}{\dd\eta}.
 \label{eq:policy-total}
\end{align}
The return derivative in the last expression is
\[
 \frac{\dd\log(V_i/C)}{\dd\eta}
 =\phi_i-\partial_\eta\log C
       -(h_i+\partial_u\log C)\frac{\dd u}{\dd\eta}.
\]
The implicit-function theorem applies when $\partial_u M\ne0$; on the decreasing branch it gives Lemma~\ref{thm:policy}. Since $\CS=u+s$ and $s$ is fixed, its consumer-surplus derivative equals $\dd u/\dd\eta$. Taking $Y_i=p_i$ and transaction weights in~\eqref{eq:normalized-expectation} also proves~\eqref{eq:transaction-price-decomposition}. The same calculations give the one-sided identities wherever the corresponding one-sided derivatives exist at advertising entry or exit.

\section{Proofs for price rewards}\label{px:appendix}

\subsection{Proof of Proposition~\ref{prop:px-general}}
\label{px:general-distributions}

\begin{proof}
Lemma~\ref{lem:conditional-price} gives the conditional prices. For the primitives in~\eqref{eq:px-general-primitive-class}, the proposed equilibrium at $\beta=0$ has
\[
 p_i=c+1/h_i,\qquad u=v_L-c+x_L-1/h_L.
\]
The gap equation makes these prices consistent with both chosen cutoffs. The type masses give visit shares $q$ and $1-q$, and the search-cost equation gives search indifference. To check entry, the low-quality type's individual search root is $v_L-c+x_s-1/h(x_s)>0$. The high-quality root is larger. Thus the candidate interval contains only positive reservation utilities and excludes all participation-boundary outcomes.

At any proposed reservation utility, let $A_i=h_i^2+h_i'$ abbreviate the denominator in the conditional price derivative, with all quantities evaluated at the corresponding conditional cutoff. Visit shares at $\beta=0$ do not depend on that utility. Hence
\[
 -M_u=\E_q[S_i h_i^2/A_i]>0.
\]
This gives a unique complete equilibrium. At the constructed equilibrium,~\eqref{eq:policy-gain-main} gives
\[
 M_\beta=\E_q[S_i/A_i]
 -q(1-q)(G_H-G_L)(1/h_H-1/h_L).
\]
By~\eqref{eq:application-price-response}, both prices rise if and only if
\begin{equation}
 -\frac{\dd u}{\dd\beta}>\max_i\frac{1}{h_i'}.
 \label{eq:px-general-price-test}
\end{equation}
Substituting $\dd u/\dd\beta=M_\beta/(-M_u)$ gives exactly~\eqref{eq:px-general-primitive-inequality}; it also gives a strict fall in consumer surplus.

The individual search roots bound all equilibria in a compact positive interval. On that interval the conditional solutions and their first derivatives are continuous, and $-M_u$ at $\beta=0$ is bounded away from zero. These properties persist on a sufficiently small positive price-weight interval and under small admissible changes in gains, hazards, hazard slopes, qualities, costs, and masses. The implicit-function theorem preserves the strict response signs there. Integrating the derivatives gives the finite comparisons.

For positive advertising, use the technology~\eqref{eq:px-weak-advertising} constructed below. On the compact conditional domain, define $V_i=(p_i-c)e^{kv_i-\beta p_i}S_i$ and $C_0=\sum_i\alpha_i e^{kv_i-\beta p_i}S_i$, and choose $K>2\max_{i,u,\beta}(C_0/V_i)$. Equations~\eqref{eq:px-advertising-solution}--\eqref{eq:px-advertising-scale} follow from the advertising and scale equations for any common match distribution. Both types advertise when $\epsilon K<1$. As $\epsilon\downarrow0$, their expenditures and multiplier responses vanish, and the conditional search-gain function converges together with its first derivatives to the no-advertising function uniformly on this domain. Its strict decrease and the strict price and consumer-surplus responses therefore persist. The unchanged positive individual roots exclude participation-boundary outcomes. This proves the advertising claim.

For the distribution examples stated in the text, logistic matches with $q=1/2$ and $h_H+h_L=1$ reduce~\eqref{eq:px-general-primitive-inequality} to~\eqref{px:main-rejection-test}. For standard-normal matches at $x_H=-1.5$, $x_L=1.5$, and $q=1/2$, the two sides of the primitive inequality are approximately $2.5085$ and $2.1785$.
\end{proof}

\subsection{Proof of Proposition~\ref{px:proposition}, parts 1 and 2}

\begin{proof}
Write $\Delta=v_H-v_L\geq0$. For logistic matches, $h(x)=F(x)$, $1-F(x)=1-h(x)$, $h'(x)=h(x)[1-h(x)]$, and $G(x)=-\log h(x)$. The entry condition in the text is therefore
\begin{equation}
 v_L-c>e^s-\log\frac{e^{-s}}{1-e^{-s}}.
 \label{px:entry}
\end{equation}

\paragraph{Conditional choices and market uniqueness.}
By Lemma~\ref{lem:conditional-price}, the conditional price is the unique global optimum. Its rejection probability $h_i=h(u+p_i-v_i)$ solves
\begin{equation}
 \log\frac{h_i}{1-h_i}-\frac{1}{h_i+\beta}=u+c-v_i,
 \qquad p_i=c+\frac{1}{h_i+\beta}.
 \label{px:conditional-h}
\end{equation}
The left side is strictly increasing from minus infinity to plus infinity. Conditional choices are therefore smooth, including at $\beta=0$.

For calculations within this appendix put
\[
 A_i=(h_i+\beta)^2+h_i(1-h_i)=\beta^2+(1+2\beta)h_i.
\]
Here $A_i$ is the derivative denominator in the price condition. Direct differentiation yields
\begin{equation}
 \partial_u p_i=-\frac{h_i(1-h_i)}{A_i},\quad \partial_\beta p_i=-\frac1{A_i},\quad
 \partial_u h_i=h_i(1-h_i)(1+\partial_u p_i),\quad \partial_\beta h_i=\partial_u p_i.
 \label{px:conditional-derivatives}
\end{equation}
In particular, $0<(-\partial_u p_i)\leq1-h_i<1$ for $\beta\geq0$. Conditional gains $G_i=-\log h_i$ strictly decrease in $u$. Their individual roots $G_i=s$ occur at
\begin{equation}
 u_i^s(\beta)=v_i-c+\log\frac{e^{-s}}{1-e^{-s}}-
                         \frac1{e^{-s}+\beta}.
 \label{px:individual-roots}
\end{equation}
Condition~\eqref{px:entry} makes both roots positive. Hence every type's gain at $u=0$ exceeds $s$, regardless of visit shares. With no advertising, conditional prices and shares at $u=0$ do not depend on consumers' boundary continuation probability. All stopping and participation-boundary equilibria are therefore excluded. Every active root lies between the two roots in~\eqref{px:individual-roots}.

The conditional visit shares and search gain are
\[
 \alpha_i\rho_i=\frac{e^{-\beta p_i}}{e^{-\beta p_H}+e^{-\beta p_L}},
 \qquad M(u,\beta)=\sum_i\alpha_i\rho_i(-\log h_i).
\]
The common response formulas specialize to
\begin{align}
 M_u&=-\mathbb E_{\rho}[(1-h_i)(1+\partial_u p_i)]
             +\beta\operatorname{Cov}_{\rho}(-\log h_i,(-\partial_u p_i)),
 \label{px:Mu}\\
 M_\beta&=\mathbb E_{\rho}\frac{1-h_i}{A_i}
       +\operatorname{Cov}_{\rho}\!\left(-\log h_i,-p_i+\frac{\beta}{A_i}\right).
 \label{px:Mbeta}
\end{align}
These are partial derivatives after conditional repricing. 

There is a useful primitive interval of global uniqueness. Let $a_0>0$ solve
\[
 \log\frac{a_0}{1-a_0}-\frac1{a_0}=c-v_H.
\]
It is the high-quality type's conditional rejection probability at $u=\beta=0$. The ordering of~\eqref{px:conditional-h} and~\eqref{px:conditional-derivatives} give $h_i(u,\beta)\geq a_0-\beta$ for all $u\geq0$. Also $G_H-G_L\leq\Delta$: holding $u,\beta$ fixed, differentiation with respect to quality gives $G_v=(1-h)(1+\partial_u p)\leq1$. Since $(-\partial_u p_i)\leq1-h_i$,
\[
 \operatorname{Cov}_{\rho}(G_i,(-\partial_u p_i))
 \leq (G_H-G_L)\mathbb E_{\rho}(-\partial_u p_i)
 \leq\Delta\mathbb E_{\rho}(1-h_i).
\]
Thus~\eqref{px:Mu} implies
\begin{equation}
 M_u\leq[-a_0+\beta(1+\Delta)]\mathbb E_{\rho}(1-h_i)<0
 \quad\text{if}\quad
 0\leq\beta\leq\frac{a_0}{2(1+\Delta)}.
 \label{px:global-unique}
\end{equation}
Search gain tends to zero as $u\to\infty$. Strict entry, strict decrease, and the intermediate value theorem establish exactly one active root, with no other consumer branch. Conditional price uniqueness then gives a unique complete equilibrium on a positive price-weight interval.

\paragraph{Classification at a price-neutral rule.}

At $\beta=0$, visit shares are $1/2$. For the classification below, rejection probabilities and conditional gain derivatives are evaluated at that equilibrium. Search indifference and the difference between the two equations in~\eqref{px:conditional-h} give
\begin{equation}
 h_H h_L=e^{-2s},\qquad
 \Delta=\frac1{h_H}-\frac1{h_L}+\log\frac{h_L(1-h_H)}{h_H(1-h_L)}.
 \label{px:gap-map}
\end{equation}
Consequently each quality gap determines exactly one pair $0<h_H\leq h_L<1$. Parameterize the rejection probabilities by their distance from the common log mean:
\begin{equation}
 h_H=e^{-s-t},\qquad h_L=e^{-s+t},\qquad 0\leq t<s.
 \label{px:q}
\end{equation}
The coordinate $t$ measures half the log ratio of the two rejection probabilities. Both terms on the right side of~\eqref{px:gap-map} strictly increase in $t$, from zero to infinity. The equilibrium reservation utility is
\[
 u=v_L-c+\log\frac{h_L}{1-h_L}-\frac1{h_L}>0.
\]

The two conditional gain derivatives at the neutral rule are
\begin{equation}
 (-M_u)=\frac{h_H(1-h_H)+h_L(1-h_L)}2>0,
 \qquad M_\beta=e^s(\cosh t-t\sinh t)-1.
 \label{px:LN}
\end{equation}
Indeed, $A_i=h_i$ and $(-\partial_u p_i)=1-h_i$ there, and~\eqref{px:Mbeta} gives
\[
 M_\beta=\frac12\left(\frac{1-h_H}{h_H}+\frac{1-h_L}{h_L}\right)
       -\frac14\log\frac{h_L}{h_H}\left(\frac1{h_H}-\frac1{h_L}\right).
\]
Implicit differentiation now yields the exact formulas
\begin{equation}
 \frac{\dd u}{\dd\beta}=\frac{M_\beta}{-M_u},\qquad
 \frac{\dd p_i}{\dd\beta}=-\frac1{h_i}-(1-h_i)\frac{M_\beta}{-M_u}
          =-\frac{(-M_u)+h_i(1-h_i)M_\beta}{h_i(-M_u)}.
 \label{px:exact-responses}
\end{equation}
The consumer-surplus response has the sign of $M_\beta$, since $s$ is fixed. A type's price rises, is unchanged, or falls according as
\begin{equation}
 M_\beta<-\frac{(-M_u)}{h_i(1-h_i)},\qquad
 M_\beta=-\frac{(-M_u)}{h_i(1-h_i)},\qquad
 M_\beta>-\frac{(-M_u)}{h_i(1-h_i)}.
 \label{px:individual-test}
\end{equation}
In particular, both prices rise if and only if
\begin{equation}
 M_\beta<-\frac{(-M_u)}{\min\{h_H(1-h_H),h_L(1-h_L)\}}.
 \label{px:both-test}
\end{equation}
Equations~\eqref{px:gap-map}--\eqref{px:both-test} classify the responses at each search cost and quality gap.

The response $M_\beta$ falls with the gap parameter $t$: for $t>0$, \mbox{$\frac{\dd M_\beta}{\dd t}=-e^s t\cosh t<0$}. Its endpoint values are
\[
 \left.M_\beta\right|_{t=0}=e^s-1>0,\qquad
 \lim_{t\uparrow s}M_\beta=\frac{(1-s)(e^{2s}-1)}2.
\]
If $s\leq1$, $M_\beta$ is positive at every finite gap, proving part 1 of Proposition~\ref{px:proposition} in that case. If $s>1$, there is exactly one zero of $M_\beta$ in $(0,s)$, and strict monotonicity of the gap map gives exactly one consumer-surplus threshold. The price tests contain additional terms, so this single-crossing statement does not apply automatically to each price. In fact, as $t\uparrow s$, $h_L\to1$ and $M_\beta$ has a finite limit;~\eqref{px:exact-responses} gives $\frac{\dd p_L}{\dd\beta}\to-1$. The low-quality price therefore falls at sufficiently large gaps even when consumer surplus falls.

\paragraph{A reversal interval and finite policy comparisons.}\label{px:reversal-region}

Consider the quality gap at which $h_H+h_L=1$. It exists when $s>\log2$. At this gap, both $h_H(1-h_H)$ and $h_L(1-h_L)$ equal $h_H h_L$, so~\eqref{px:both-test} reduces exactly to
\[
 (h_L-h_H)\log(h_L/h_H)>2.
\]
Under~\eqref{px:q}, the restriction $h_H+h_L=1$ is $s=\log(2\cosh t)$, and the last inequality is $t\tanh t>1$. The function $t\tanh t$ increases strictly from zero to infinity. Let $t_*>0$ be its unique unit crossing, and define
\begin{equation}
 t_*\tanh t_*=1,\qquad
 s_*=\log(2\cosh t_*)\simeq1.2865682649.
 \label{px:threshold}
\end{equation}
For every $s>s_*$, the gap
\begin{equation}
 \Delta_s=e^{2s}\sqrt{1-4e^{-2s}}
                +4\operatorname{arcosh}(e^s/2)
 \label{px:reference-gap}
\end{equation}
therefore produces strict price increases at $\beta=0$. This search-cost threshold is sufficient for a reversal region; it is not asserted to be the smallest search cost at which any quality gap can produce one.

To turn the local reversal into a finite policy comparison, we need the response signs and uniqueness bound to hold uniformly on one positive policy interval. Fix $s>s_*$ and a low quality satisfying~\eqref{px:entry}. The strict response inequalities at $\Delta_s$, together with the continuous gap map, permit a closed interval $J$ of positive length around $\Delta_s$ on which both price derivatives are strictly positive at $\beta=0$. Their minima on $J$ are positive. The corresponding consumer derivative is strictly negative throughout $J$. The welfare derivative is also strictly negative by the calculation below.

As $\Delta$ varies over $J$, the number $a_0$ in~\eqref{px:global-unique} has a positive minimum. Choose an initial $B>0$ below the minimum of $a_0/[2(1+\Delta)]$ over $J$. All equilibria for $(\Delta,\beta)\in J\times[0,B]$ are unique and active. The individual roots in~\eqref{px:individual-roots} place them in a common compact positive interval. On this interval all conditional solutions are smooth, rejection probabilities are bounded away from zero and one, and~\eqref{px:global-unique} bounds $M_u$ away from zero. The implicit function theorem therefore gives continuous equilibrium derivatives jointly in $\Delta,\beta$ on the rectangle, including their right limits at zero. Uniform continuity preserves the positive minima of both price derivatives and the negative maxima of both surplus derivatives after reducing $B$ to a positive $\bar\beta$. This reduction does not affect global uniqueness or entry. Integrating each strict derivative over any subinterval proves the finite policy comparisons. This proves part 2 of Proposition~\ref{px:proposition}.

The same compact argument at $\Delta=0$ yields a small-gap interval and a positive price-weight interval with both price derivatives negative and the consumer derivative positive. At fixed $s$, the reversal comparison thus changes the quality gap while holding the rule and all other primitives fixed.

\paragraph{Total surplus.}\label{px:welfare}

Use transaction shares
\[
 \omega_i=\frac{e^{-\beta p_i}(1-h_i)}
                   {\sum_j e^{-\beta p_j}(1-h_j)}.
\]
With no advertising, industry profit is $\E_\omega[p_i-c]$ and total surplus is $W=u+s+\E_\omega[p_i-c]$. Partial derivatives in the next display hold $u$ fixed or hold $\beta$ fixed, as indicated. Differentiating the normalized transaction weights, using~\eqref{px:conditional-derivatives} and $\beta+h_i=1/(p_i-c)$, gives
\begin{align}
 \left.\frac{\partial W}{\partial\beta}\right|_u
   &=-\operatorname{Var}_\omega(p_i-c)
      -\E_\omega[p_i-c]\,\mathbb E_\omega\frac1{A_i(p_i-c)}<0,
       \label{px:welfare-fixed}\\
 \left.\frac{\partial W}{\partial u}\right|_\beta
   &=\E_\omega[p_i-c]\,\mathbb E_\omega\frac{1+\partial_u p_i}{(p_i-c)}>0.
       \label{px:welfare-u}
\end{align}
For completeness, the conditional log transaction-weight derivatives before normalization are $-p_i+1/(A_i(p_i-c))$ and $\beta-(1+\partial_u p_i)/(p_i-c)$. The derivative of an expectation equals the expectation of the direct derivative plus its covariance with this log-weight derivative. Thus the first welfare derivative is
\[
 -\mathbb E_\omega\frac1{A_i}
 +\operatorname{Cov}_\omega\left((p_i-c),-(p_i-c)+
                                      \frac1{A_i(p_i-c)}\right),
\]
which gives~\eqref{px:welfare-fixed}. The second is
\[
 1-\mathbb E_\omega (-\partial_u p_i)+
 \operatorname{Cov}_\omega\left((p_i-c),-\frac{1+\partial_u p_i}{(p_i-c)}\right),
\]
which gives~\eqref{px:welfare-u}. Therefore $\frac{\dd u}{\dd\beta}\leq0$ implies $\frac{\dd W}{\dd\beta}<0$ at every regular active equilibrium, including positive price weights. Both-price reversals require $\frac{\dd u}{\dd\beta}<0$ by the price response formula. Their consumer and total-surplus losses follow.\qedhere

\end{proof}

\subsection{Proof of the quality-weight comparison in Section~\ref{sec:prices}}
\label{px:positive-quality-weight}\label{lem:px-quality-necessary}

Retain the two-type logistic market, common marginal cost, and equal type
masses. In this subsection advertising is absent, but the score is
$\exp(kv_i-\beta p_i)$ for a fixed $k\geq0$. Conditional prices and the
entry bound~\eqref{px:entry} do not depend on $k$. A quality weight multiplies
the return at every price by the same factor. The conditional derivative
identities in~\eqref{px:conditional-derivatives} therefore continue to hold.
Visit shares now satisfy
\[
 \alpha_i\rho_i=
 \frac{\exp(kv_i-\beta p_i)}
 {\exp(kv_H-\beta p_H)+\exp(kv_L-\beta p_L)}.
\]
Equations~\eqref{px:Mu}--\eqref{px:Mbeta} remain valid with these shares.
The proof of~\eqref{px:global-unique} uses bounds on conditional hazards and
on the range of gains, rather than equal visit shares. It consequently
establishes the same positive interval of global uniqueness for every fixed
$k\geq0$.

\begin{proof}
Write $\Delta=v_H-v_L$. At the price-neutral rule, the high-quality visit
share is
\[
 R=\frac{e^{k\Delta}}{1+e^{k\Delta}}.
\]
It is independent of the proposed reservation utility. Search indifference
therefore gives
$R\log h_H+(1-R)\log h_L=-s$. If $\Delta>0$, put
$t=\log(h_L/h_H)>0$. These rejection probabilities can be written as
\begin{equation}
 h_H=e^{-s-(1-R)t},\qquad h_L=e^{-s+Rt},\qquad 0<t<s/R.
 \label{eq:px-weighted-hazards}
\end{equation}
The upper bound follows from $h_L<1$. For $\Delta=0$, the same expressions
hold with $t=0$.

Specializing the conditional response identities at $\beta=0$ yields
\begin{align}
 -M_u={}&R h_H(1-h_H)+(1-R)h_L(1-h_L)>0,\notag\\
 M_\beta={}&R\frac{1-h_H}{h_H}+(1-R)\frac{1-h_L}{h_L}
 -R(1-R)t\left(\frac1{h_H}-\frac1{h_L}\right).
 \label{eq:px-weighted-responses}
\end{align}
Substitution from~\eqref{eq:px-weighted-hazards} gives
\[
 M_\beta=e^s\left\{
 R e^{(1-R)t}+(1-R)e^{-Rt}
 -R(1-R)t\bigl[e^{(1-R)t}-e^{-Rt}\bigr]
 \right\}-1.
\]
For an algebraic lower bound, hold $R$ fixed and vary $t$. The derivative
of this expression is
\[
 -e^sR(1-R)t\bigl[(1-R)e^{(1-R)t}+Re^{-Rt}\bigr]<0
 \qquad(t>0).
\]
Its limit as $t$ tends upward to $s/R$ equals
\[
 [R-(1-R)s](e^{s/R}-1).
\]
At every finite quality gap, $t<s/R$. Thus $s\leq R/(1-R)$ implies
$M_\beta>0$. Since $R/(1-R)=e^{k\Delta}$ and
$du/d\beta=M_\beta/(-M_u)$, this proves the consumer-surplus statement.
The price response is
\[
 \frac{dp_i}{d\beta}
 =-\frac1{h_i}-(1-h_i)\frac{du}{d\beta}.
\]
It is strictly negative when $du/d\beta\geq0$, and either price can rise
only when $du/d\beta<0$. This proves the remaining claims.
To verify the large-gap comparison, fix $k>0$ and $s$, and let the quality
gap increase without bound. Then $R\to1$. The conditional pricing equations
and search indifference imply $h_H\to e^{-s}$ and $h_L\to1$; hence
\eqref{eq:px-weighted-responses} gives $M_\beta\to e^s-1>0$.
This proves the eventual consumer benefit.
\end{proof}

\subsection{Proof of Proposition~\ref{px:proposition}, part 3}
\label{px:positive-advertising}

\begin{proof}
We use a common advertising technology
\begin{equation}
 g_\epsilon(z)=1+\epsilon K(1-e^{-z/\epsilon}),\qquad z\geq0,
 \label{eq:px-weak-advertising}
\end{equation}
where $K>0$ fixes the initial marginal return and $\epsilon>0$ controls the size of the multiplier and its response. The function is strictly increasing and strictly concave. We will choose $K$ so both types advertise and then make $\epsilon$ small enough to preserve the strict inequalities.

Fix $s>s_*$, common cost $c$, and low quality satisfying~\eqref{px:entry}. The strict no-advertising reversal at $\Delta_s$ in
\eqref{px:reference-gap}, the global bound~\eqref{px:global-unique}, and
continuity give a closed gap interval $J$ of positive length and positive
$\bar\beta,\bar k$ on which both price derivatives are uniformly positive and the consumer- and total-surplus derivatives uniformly negative. Shrink these bounds if needed. The single-type roots in
\eqref{px:individual-roots} bracket all equilibria in one compact subset
of $(0,\infty)$, and $M_u$ is uniformly negative there. These properties
continue to hold for $k\in[0,\bar k]$ by the preceding subsection.

On that compact conditional domain, put
$V_i=(p_i-c)e^{kv_i-\beta p_i}S_i$ and
$C_0=\sum_i\alpha_i e^{kv_i-\beta p_i}S_i$.
Choose $K>2\max_{i,u,\Delta,k,\beta}(C_0/V_i)$ and
$\epsilon K<1$. Under~\eqref{eq:px-weak-advertising}, the interior
advertising conditions and scale consistency have the explicit solution
\begin{align}
 z_i&=\epsilon\log(KV_i/C),\qquad
 g_\epsilon(z_i)=1+\epsilon K-\epsilon C/V_i,
 \label{eq:px-advertising-solution}\\
 C&=\frac{(1+\epsilon K)C_0}
          {1+\epsilon\sum_i\alpha_i/(p_i-c)}.
 \label{eq:px-advertising-scale}
\end{align}
Because $C<2C_0<KV_i$, both advertising expenditures are positive. Strict concavity
verifies global advertising optimality, and conditional uniqueness verifies
the scale.

These solutions and their first derivatives in $u,k,\beta$
converge uniformly to the no-advertising choices as $\epsilon\downarrow0$;
in particular, $z_i=O(\epsilon)$,
$0<\chi_i\leq\epsilon K$, where $\chi_i$ is the proportional multiplier response defined in~\eqref{eq:advertising-response}.
The conditional gain function and its first derivatives consequently
converge uniformly. Its strict decrease on the entire candidate root
interval survives for sufficiently small $\epsilon$.

Advertising does not change the single-type gain roots or conditional
prices. Their positive bracket and~\eqref{px:entry} exclude every stopping
or participation-boundary outcome independently of visit shares. The active
root is therefore globally unique. The implicit-function theorem gives
uniform convergence of equilibrium responses. Applying the same argument
to $W=u+s+\E_\omega[p_i-c]-\sum_i\alpha_i z_i$ preserves its strict
response as well. Choose $\bar\epsilon>0$ small enough that every $\epsilon\in(0,\bar\epsilon]$ preserves
the price, consumer, and welfare signs throughout the parameter rectangle;
integration gives the finite comparisons.

Finally, after any allowed advertising history, each type's aggregate
advertising multiplier is a fixed positive constant during its price and
search continuation. The conditional score derivative in $u$ is then
$-\beta\partial_u p_i$. The share-independent bound
\eqref{px:global-unique}, the unchanged single-type bracket, and strict
entry again give a unique continuation. This also permits different
advertising expenditures among firms of the same type.
\end{proof}

\section{Proofs of the product-guidance results}
\label{app:quality}

\subsection{Proof of Theorem~\ref{thm:quality}}
\label{app:quality-finite}

\begin{proof}
We establish the conditional comparison, then order the equilibrium extrema.

Compare quality weights $k_1>k_0\geq0$, using superscripts $j=0,1$ for the two rules. Fix a candidate consumer state $(u,\vartheta)$ and order types by quality. The conditional pricing equation is $1/(p_i-c)=h(u+p_i-v_i)+\beta$. Consider a quality increment $\delta>0$. At any fixed price it weakly lowers the right side, so the unique conditional price weakly increases. That price increase is strictly smaller than $\delta$: an increase of at least $\delta$ would strictly lower the inverse markup while weakly raising the acceptance cutoff and its hazard, contradicting the pricing equation. The acceptance cutoff therefore falls strictly with quality. This argument uses monotonicity and does not differentiate the hazard.

Under a nondecreasing hazard, $G(x)\leq S(x)/h(x)$, obtained by integrating $S(x+y)\leq S(x)e^{-h(x)y}$. Hence $(G/S)'= -1+hG/S\leq0$. Both $S_i$ and $G_i/S_i$ are nondecreasing in quality. The multiplier ratio $e^{(k_1-k_0)v_i}$ has the same ordering. Conditional prices are identical under the two rules: the characteristic multiplier and the advertising expenditure factor multiply operating returns without changing their price maximizer.

The continuation probability $\vartheta\in[0,1]$ matters for the common demand scale. Put $d_i=1-\vartheta+\vartheta S_i$, $C^j=\sum_i\alpha_iw_i^jd_i$, and $\pi_i^j=\alpha_iw_i^jd_i/C^j$, the normalized probability that the terminating visit is of type $i$ under rule $j$. Advertising expenditure maximizes $(e^{k_jv_i-\beta p_i}(p_i-c)S_i/C^j)g_i(z)-z$. Its unique optimizer is nondecreasing in the coefficient, including at zero. Hence, with $q_i=(e^{(k_1-k_0)v_i})/(C^1/C^0)$,
\begin{equation}
 \frac{\pi_i^1}{\pi_i^0}
 =q_i\frac{g_i(z_i^1)}{g_i(z_i^0)},\qquad
 \operatorname{sgn}(\pi_i^1-\pi_i^0)=\operatorname{sgn}(q_i-1).
 \label{eq:app-quality-singlecross}
\end{equation}
The differences $\Delta_i=\pi_i^1-\pi_i^0$ therefore change sign at most once, from negative to positive, and sum to zero. If all vanish the comparison is immediate. Otherwise, for every nondecreasing test sequence $Y_i$, choose $Y_*$ between its values on the two sides of the crossing. Then $\sum_i\Delta_iY_i=\sum_i\Delta_i(Y_i-Y_*)\geq0$.

The opportunity ordering makes $1/d_i$ nonincreasing and $G_i/d_i=(G_i/S_i)S_i/(1-\vartheta+\vartheta S_i)$ nondecreasing. Let $M^j(u,\vartheta)$ denote the conditional expected search gain under rule $j$. Its representation below has a numerator that rises and a positive denominator that falls:
\begin{equation}
 M^j(u,\vartheta)
 =\frac{\sum_i\pi_i^jG_i/d_i}{\sum_i\pi_i^j/d_i},
 \qquad M^1(u,\vartheta)\geq M^0(u,\vartheta).
 \label{eq:app-quality-ratio}
\end{equation}
This finite comparison requires no derivatives at advertising entry. At active search, $\vartheta=1$, $d_i=S_i$, and $\pi_i^j=\omega_i^j$, giving the main-text representation~\eqref{eq:quality-transaction-search-gain}.

To pass from conditional gains to equilibrium surplus, consider each consumer branch. These are stopping at $(u,\vartheta)=(0,0)$ when $M^j(0,0)\leq s$, a boundary state with $M^j(0,\vartheta)=s$, and active search with $u>0$, $\vartheta=1$, $M^j(u,1)=s$. Their surplus values are respectively $M^j(0,0)$, $s$, and $s+u$. At $\vartheta=0$, consumers stop after the free first visit; at $\vartheta=1$ they continue whenever the offer is unacceptable. Intermediate values are possible only at indifference. The price optimizer depends on $u$ but not on $\vartheta$, so all zero-utility states give the same conditional prices. Equilibrium exists by Theorem~\ref{thm:aggregation-en}. Continuity closes the branch intersections, and $M^j(u,1)\leq\max_iG_i(u-v_i+c_i)\to0$ bounds active roots. Thus equilibrium extrema exist.

If the old minimum surplus is below $s$, it is stopping surplus; the new minimum is either its weakly larger stopping surplus or at least $s$. If the old minimum equals $s$, then $M^0(0,0)\geq s$, excluding a new stopping outcome below $s$. If the old minimum exceeds $s$, every old boundary value satisfies $M^0(0,\vartheta)>s$: otherwise stopping or an intermediate boundary root would exist. Before the smallest old active root $L$, continuity also gives $M^0(u,1)>s$. Both inequalities persist under the new policy, so its smallest root is at least $L$.

If the largest old reservation utility is $U>0$, then $M^1(U,1)\geq s$ and the vanishing tail supplies a new root at or above $U$. If the old maximum surplus is $s$, some old boundary value equals $s$. Its new value is at least $s$; either $M^1(0,0)>s$, so every new outcome has surplus at least $s$, or continuity supplies a new boundary root. Finally, an old maximum below $s$ is stopping surplus, and every new outcome is either improved stopping surplus or at least $s$. These arguments establish both surplus extrema.

They also handle the possible switch from stopping immediately to active search: a positive reservation utility always yields surplus above $s$, whereas a zero-utility outcome never does. Thus whenever an old reservation-utility extremum is positive the corresponding reservation-utility comparison follows from its surplus comparison. When the old minimum reservation utility is zero, the new minimum cannot be lower; when the old maximum reservation utility is zero the new maximum likewise cannot be lower. Since each $p_i(u)$ is policy independent and nonincreasing, its largest equilibrium price is attained at the smallest reservation utility and its smallest price at the largest reservation utility. Both price extrema therefore decrease weakly.

\end{proof}

\subsection{Proof of Corollary~\ref{cor:quality-selection}}
\label{app:quality-selection}

\begin{proof}
Write $a=\alpha_H\in(0,1)$ for the favored population's firm mass,
$\delta=v_H-v_L>0$ for the quality gap, and
$\bar p=\E_\omega p_i$ for the average transaction price. Fix $\beta\geq0$.
Define $m(x)=1/[\beta+h(x)]$; primes in this proof denote derivatives with
respect to the required match $x$. Conditional pricing can be written as
\begin{equation}
 x_i-m(x_i)=u+c-v_i,\qquad p_i=c+m(x_i),
 \label{eq:selection-conditional-cutoff}
\end{equation}
where $x_i=u+p_i-v_i$. The left side is strictly increasing because the
hazard is nondecreasing.

\paragraph{A unique active market with positive advertising.}
Let $x_s$ solve $G(x_s)=s$. The single-quality reservation utilities are
$u_L^s=v_L-c+x_s-m(x_s)$ and $u_H^s=u_L^s+\delta$.
Every active equilibrium belongs to $[u_L^s,u_H^s]$: below this interval both
conditional search gains exceed $s$, and above it both are smaller than $s$.
The assumed $u_L^s>0$ excludes a participation boundary, because both
conditional gains at $u=0$ exceed $s$, irrespective of visit shares.
At equal qualities, every firm has demand one and advertising expenditure
$\gamma m(x_s)-\kappa>0$. Continuity of the unique conditional scale and
choices keeps both types advertising on the shrinking candidate interval,
uniformly over the prescribed compact set of quality weights. At equal
qualities,
\[
 M_u=-\frac{S(x_s)}{1-m'(x_s)}<0,
\]
where $S(x)=1-F(x)$ is acceptance probability. Smoothness and continuity
preserve this strict inequality on the whole candidate interval for small
$\delta$. The search equation therefore has one root. Conditional price and
scale uniqueness give a unique complete active equilibrium.

\paragraph{Transaction prices at prescribed visit shares.}
We first isolate purchase selection by prescribing the composition of visits.
Prescribe the high-quality visit share $r$ near $a$, and use $\E_r$ for the
visit mixture $(1-r,r)$. The equations~\eqref{eq:selection-conditional-cutoff}
and $\E_rG(x_i)=s$ have a smooth local solution. Taylor expansion and the
implicit-function theorem give, uniformly on a compact neighborhood,
\[
 u=u_L^s+r\delta+O(\delta^2),\qquad
 x_i-x_s=\frac{r\delta-(v_i-v_L)}{1-m'(x_s)}+O(\delta^2).
\]
Because $G'=-S$ and $G''=hS$, search indifference implies
\[
 \E_r(x_i-x_s)=\frac{h(x_s)}2\E_r[(x_i-x_s)^2]+O(\delta^3).
\]
Expanding $\bar p-c=\E_r[S(x_i)m(x_i)]/\E_rS(x_i)$ consequently yields
\begin{align}
 \bar p(r,\delta)
 &=c+m(x_s)+B\,r(1-r)\delta^2+O(\delta^3),
 \label{eq:selection-pbar-expansion}\\
 B&=\frac{m''(x_s)-h(x_s)m'(x_s)}{2[1-m'(x_s)]^2}.
 \label{eq:selection-variance-coefficient}
\end{align}
The remainder and its first $r$-derivative are $O(\delta^3)$; four continuous
hazard derivatives suffice for these uniform bounds. The fixed-share utility
also satisfies $u_r=\delta+O(\delta^2)$. Direct differentiation of $m$ gives
\begin{equation}
 \operatorname{sgn} B
 =\operatorname{sgn}\!\left\{2[h'(x_s)]^2+
 [\beta+h(x_s)]\bigl[h(x_s)h'(x_s)-h''(x_s)\bigr]\right\}.
 \label{eq:selection-B-primitives}
\end{equation}
Thus the curvature condition in the corollary is expressed entirely in the
match distribution, search cost, price weight, and firm mass.

\paragraph{How quality guidance changes composition.}
The next step determines how the recommendation rule changes the visit shares used in the preceding expansion. The interior advertising condition is
$z_i=\gamma(p_i-c)D_i-\kappa$. Combining it with
$D_i=w_iS(x_i)/C$ eliminates the common scale from relative scores:
\begin{equation}
 \frac r{1-r}=\frac a{1-a}
 \exp\!\left(\frac{k\delta-\beta(p_H-p_L)}{1-\gamma}\right)
 \left[\frac{(p_H-c)S(x_H)}{(p_L-c)S(x_L)}\right]^{\gamma/(1-\gamma)}.
 \label{eq:selection-share-equation}
\end{equation}
Prices and cutoffs on the right are the prescribed-share outcomes above.
At $\delta=0$, the equation has solution $r=a$, and its log residual has
$r$-derivative $1/[a(1-a)]$. The smooth solution has $r=a+O(\delta)$.
Only $k\delta/(1-\gamma)$ depends directly on $k$, giving
\begin{align*}
 \frac{\dd r}{\dd k}&=\frac{a(1-a)}{1-\gamma}\delta+O(\delta^2),\\
 \frac{\dd u}{\dd k}&=\frac{a(1-a)}{1-\gamma}\delta^2+O(\delta^3)>0,\\
 \frac{\dd\bar p}{\dd k}
 &=B\frac{a(1-a)(1-2a)}{1-\gamma}\delta^3+O(\delta^4)>0.
\end{align*}
All remainders are uniform over the fixed compact policy interval.
The last sign follows from $(1-2a)B>0$. Consumer surplus rises since search
cost is fixed. Conditional prices have no direct $k$ dependence, and
\[
 \frac{\dd p_i}{\dd k}=\frac{\partial p_i}{\partial u}
 \frac{\dd u}{\dd k}<0,\qquad
 \frac{\partial p_i}{\partial u}
 =-\frac{h'(x_i)}{[\beta+h(x_i)]^2+h'(x_i)}<0.
\]
The local positivity of $h'$ ensures the strict price cuts.
Aggregating the advertising condition gives
\eqref{eq:quality-advertising-accounting}, so
$\dd\Ads/\dd k=\gamma\,\dd\bar p/\dd k>0$.
Uniform strict signs integrate over any two distinct quality weights in the
prescribed interval.

\paragraph{Verification of the distributional conditions.} If $(1-2a)B<0$, the same argument instead gives a
fall in the average transaction price and advertising, alongside higher
consumer surplus and lower prices for every product. The case
$(1-2a)B=0$ requires higher-order terms. A sufficient condition for $B>0$
at any fixed $\beta$ is $h'(x_s)>0$ and $h''(x_s)\leq h(x_s)h'(x_s)$.
For logistic matches, put $q=e^{-s}=h(x_s)$. Then
\[
 \operatorname{sgn} B
 =\operatorname{sgn}\!\left[\frac{2q(1-q)}{\beta+q}-1+3q\right].
\]
It is positive for every finite $\beta\geq0$ if $s\leq\log3$; if
$s>\log3$, it is positive exactly when
$\beta<q(1+q)/(1-3q)$.

For a non-logistic full-support example, take standard-normal matches and
$s=1/\sqrt{2\pi}$, so $x_s=0$. With $q=h(0)=\sqrt{2/\pi}$,
$h'(0)=q^2$ and $h''(0)=2q^3-q$. The curvature numerator is
$q^4+q^2+\beta q(1-q^2)>0$ for every fixed $\beta\geq0$.
Thus a less numerous high-quality population satisfies the selection condition
whenever the baseline participation and advertising restrictions hold.
\end{proof}

\subsection{Proof of Proposition~\ref{prop:match-ad-reversal}}
\label{app:match-distributions}

\begin{proof}
All policy derivatives below hold search cost fixed. Write $q_i=\alpha_i\rho_i$ only within this proof for a type's baseline visit share, $S_i=1-F_i(u+p_i-v_i)$ for its acceptance probability, and $G_i=G_i(u+p_i-v_i)$ for its inspection gain. Thus $\sum_iq_iG_i=s$ at the baseline. The reform multiplies the score of the baseline target $H$ by $e^\eta$.

\paragraph{The general sign condition.}
At fixed reservation utility, the multiplier leaves conditional prices unchanged.
Substituting the scale and score responses from~\eqref{eq:policy-scale}
into~\eqref{eq:policy-N} gives
\begin{equation}
 \partial_\eta M
 =(1+\chi_H)q_H\left\{G_H-s-
 \frac{S_H\Cov_\rho(G_i,\chi_i)}{\E_\rho[S_i(1+\chi_i)]}\right\}.
 \label{match:general-response}
\end{equation}
With advertising expenditures frozen at the same baseline choices, $\partial_\eta M=q_H(G_H-s)$. Thus $G_H>s$ gives a positive conditional response with fixed advertising. On decreasing conditional-search branches, Lemma~\ref{thm:policy} translates the two conditional signs into the stated consumer-surplus comparison. If the hazard derivative is positive at every priced cutoff, $\partial_up_i<0$, and all prices have the opposite signs.

For two types, with $G_H>G_L$, the endogenous derivative simplifies to
\[
 \partial_\eta M
 =\frac{(1+\chi_H)(1+\chi_L)q_Hq_L(G_H-G_L)\E_\rho S_i}
 {\E_\rho[S_i(1+\chi_i)]}>0.
\]
Equal realized responses also make the covariance in~\eqref{match:general-response} zero. In particular, the common power technology cannot cause the reversal if every type advertises positively. The construction below uses three types and one strict zero-advertising choice under the common technology.

\paragraph{A primitive family.}
We first construct a strict reversal with uniform matches, then add smooth noise to obtain full-support distributions satisfying the maintained assumptions. Let all types have common quality $v$, cost $c$, and zero price weight. Type $i$ has match distribution uniform on $[-a_i,a_i]$, where $a_i>0$ is its half-width. For a proposed reservation utility put $d=v-c-u$ and assume $0<d<3\min_i a_i$. The conditional price, acceptance probability, and inspection gain are
\begin{equation}
 p_i-c=\frac{a_i+d}{2},\qquad
 S_i=\frac{a_i+d}{4a_i},\qquad
 G_i=\frac{(a_i+d)^2}{16a_i},\qquad
 \partial_up_i=-\frac12.
 \label{match:uniform-price}
\end{equation}
These prices globally maximize the conditional uniform demand objective: the displayed restriction places its maximizer strictly inside the acceptance interval. Its operating return per unit of score is $(p_i-c)S_i=2G_i$.

Let $\chi=\gamma/(1-\gamma)$ for the common technology~\eqref{eq:quality-common-advertising}. Choose positive masses $\alpha_H,\alpha_M,\alpha_L$ summing to one and a proposed positive advertising multiplier $r>1$ for $H$. Define $g_H=r$ and $g_M=r(G_M/G_H)^\chi$. If $g_M>1$ and $r(G_L/G_H)^\chi<1$, the first two types advertise positively and $L$ is strictly at zero. The corresponding original technology parameter and search cost are
\begin{align}
 C&=\alpha_HrS_H+\alpha_Mr(G_M/G_H)^\chi S_M+\alpha_LS_L,
 &\kappa&=\frac{2\gamma G_Hr^{-1/\chi}}{C},
 \label{match:primitive-family}\\
 s&=\frac{\alpha_HrG_H+\alpha_Mr(G_M/G_H)^\chi G_M+\alpha_LG_L}
 {\alpha_Hr+\alpha_Mr(G_M/G_H)^\chi+\alpha_L}.
 \label{match:primitive-searchcost}
\end{align}
The score unit is chosen so that the common zero-advertising score is one. This rescaling has no economic effect. The resulting advertising expenditures are $z_i=\kappa(g_i^{1/\gamma}-1)$ for $H,M$, and zero for $L$. They satisfy the actual scale equation and advertising optimality conditions. The common quality multiplier cancels from all these calculations.

There is a useful population-and-technology test at equal gains. Take $d=1$, $a_H=a$, $a_M=1/a$, and $a_L=1$, with $1/3<a<1$. Then
\[
 G_H=G_M=\frac{(1+a)^2}{16a}>G_L=\frac14,\qquad
 S_H=\frac{1+a}{4a}>S_M=\frac{1+a}{4},\qquad S_L=\frac12.
\]
At this tied-gain benchmark, the endogenous derivative is negative exactly when
\begin{equation}
 \chi\alpha_Mr(S_H-S_M)
 >\alpha_HrS_H+\alpha_MrS_M+\alpha_LS_L.
 \label{match:primitive-inequality}
\end{equation}
To express the condition through the advertising expenditure scale, define
\[
 K(r)=\frac{2\gamma G_Hr^{-1/\chi}}
 {r(\alpha_HS_H+\alpha_MS_M)+\alpha_LS_L}.
\]
This function is strictly decreasing and $\kappa=K(r)$. Let
\[
 r_{\max}=(G_H/G_L)^\chi,\quad
 D_\alpha=\chi\alpha_M(S_H-S_M)-\alpha_HS_H-\alpha_MS_M.
\]
If $D_\alpha>0$ and
$r_*:=\max\{1,\alpha_LS_L/D_\alpha\}<r_{\max}$, every
$K(r_{\max})<\kappa<K(r_*)$ gives strict advertising participation and~\eqref{match:primitive-inequality}. The search cost is then~\eqref{match:primitive-searchcost}. Since $r_{\max}>1$, this interval is nonempty exactly when $D_\alpha>\alpha_LS_L/r_{\max}$. Substituting the displayed acceptance probabilities and multiplying by $4a$ gives~\eqref{eq:match-width-condition}.

For a strict highest-gain target, let $G_H>G_M>G_L$, put $\Delta=G_H-G_M$, $D=G_M-G_L$, $Q=\sum_iq_iS_i$, and $T=Q+\chi(q_HS_H+q_MS_M)$. These temporary proof quantities are calculated from~\eqref{match:primitive-family}. Direct algebra in~\eqref{match:general-response} shows that its sign is negative exactly when
\[
 B:=\chi q_M(S_H-S_M)-Q>0,\qquad
 0<\Delta<\frac{q_LDB}{A},\qquad
 A:=(1-q_H)T-\chi q_Hq_LS_H>0.
\]
The last positivity follows from
$A=(q_M+q_L)Q+\chi q_M[q_HS_H+(q_M+q_L)S_M]$.
Replacing the shares by
$q_i=\alpha_i g_i/\sum_j\alpha_jg_j$ makes this an explicit inequality in the widths, masses, exponent, and multiplier in~\eqref{match:primitive-family}.

\paragraph{Strict reversal and unique implementation.}
Consider the primitive subfamily
\begin{gather*}
 v=5,\quad c=2,\quad d=1,\quad
 a_H=\frac25,\quad a_L=1,\quad \frac{249}{100}<a_M<\frac52,\\
 (\alpha_H,\alpha_M,\alpha_L)=
 \left(\frac5{27},\frac59,\frac7{27}\right),
 \quad\gamma=\frac7{10},\quad r=\frac75,
\end{gather*}
with $\kappa,s$ given by~\eqref{match:primitive-family}--
\eqref{match:primitive-searchcost}; its baseline is $u=2$.
Throughout the width interval, $G_H>G_M>G_L$, $g_M>1.39$, and
$r(G_L/G_H)^\chi<1$. Moreover,
$q_H=q_L\in[.2,.201]$, $q_M\in[.598,.6]$, $S_M<.351$, and
$G_H-G_M<.00053$. Therefore $B>.244$, $D>.0557$, $A<1.112$, and
$q_LDB/A>.00244>.00053$, proving the strict reversal uniformly.

For uniqueness, $.294<s<.296$; at $d=.9$ every gain is below $.294$,
whereas at $d=1.18$ every gain exceeds $.296$.
Uniform conditional gains are globally nondecreasing in $d$, becoming
$a_i>s$ once $d\ge3a_i$. Thus every type-specific indifference level
and every market root satisfy $.9<d<1.18$, so $u>1.82$ and participation
boundaries are excluded. Every priced cutoff on this candidate interval
is interior. Equation~\eqref{match:uniform-price} gives
$\min_iS_i\ge17/50$, $h_+\le20/13$, $d_-\ge25/92$, and
$361/1600\le G_i\le6241/16000$, hence
$\Delta_G\le2631/16000$. The power technology bounds every realized
advertising response by $7/3$, including entry points, while the price-score
range is zero. The primitive test of Appendix~\ref{app:uniqueness} gives
\begin{equation}
 \Delta_G\Delta_J
 \le\frac{2631}{16000}\frac73
 \left(\frac{20}{13}-\frac{25}{92}\right)
 =\frac{1860117}{3827200}
 <\frac{17}{25}
 \le4\min_i\{S_i(1+\partial_up_i)\}.
 \label{match:uniqueness-bound}
\end{equation}
This proves global uniqueness without prescribing which types advertise positively.
After any fixed advertising history the conditional prices and candidate
interval are unchanged; with advertising held fixed, the score-response
range is zero, giving a unique price and search continuation as well.

The uniform construction supplies strict margins for both the reversal and uniqueness. To satisfy Assumption~\ref{ass:match-en}, replace the auxiliary matches by
$\varepsilon_i=U[-a_i,a_i]+N(0,\tau^2)$, with independent common noise
and $\tau>0$. Their means remain zero; their densities are smooth,
positive, and log-concave, implying full support, a nondecreasing hazard,
and finite positive-part expectations. The candidate cutoffs are uniformly
separated from the uniform support endpoints, so survival functions,
densities, their required derivatives, and conditional prices converge
uniformly as $\tau\downarrow0$.
For any fixed interior width, every gain ranking, advertising participation
condition, reversal inequality, root bracket, and uniqueness bound has
strict slack. Uniform convergence and the implicit-function theorem
preserve them for small positive $\tau$, small positive $\beta$, and
small changes in common quality, common cost, masses, search cost, and
technology parameters. The perturbed individual indifference levels
remain inside the enlarged candidate interval. The same range test covers
every configuration of positive advertisers and every fixed advertising history, retaining only the
direct price-score term in the latter case. Price derivatives remain
strictly negative near $-1/2$, and continuity supplies a nonempty positive
reform interval with the opposite price and consumer-surplus comparisons
under adjusting and frozen advertising. This proves the claimed open
region of admissible primitives in part~(ii) of
Proposition~\ref{prop:match-ad-reversal}.
\end{proof}

\subsection{Verification of Table~\ref{tab:match-ad-comparison}}\label{app:match-illustration}
Use $\varepsilon_i=U[-a_i,a_i]+N(0,\tau^2)$ with
$(a_H,a_M,a_L)=(.4,2.41,1)$ and $\tau=.01$. Common quality and cost are $v=5,c=2$, the price weight is $\beta=.01$, and the advertising exponent is $\gamma=.7$. The remaining primitives, with displayed values rounded, are
\begin{gather*}
 (\alpha_H,\alpha_M,\alpha_L)
 =(.177957767151129,.572159185820940,.249883047027931),\\
 \kappa=.6078795902853515,\qquad s=.2962374905992597.
\end{gather*}
At the baseline $u=2$, visit shares are $(.2,.6,.2)$. Acceptance probabilities are approximately $(.8781,.3567,.5025)$ and inspection gains $(.30846,.30673,.25253)$. Thus $H$ is strictly best. Advertising expenditures are $(.37516201,.29527610,0)$; $L$'s initial marginal return to advertising is $.94014073<1$. The two positive advertising choices and the strict zero choice solve the common scale and actual firm optimality conditions.

The equations give $\dd u/\dd\eta\simeq.00985593$ with fixed advertising expenditure
and $\dd u/\dd\eta\simeq-.00154403$ with advertising expenditure adjusting. Both conditional
search derivatives are negative, approximately $-.2480$ and $-.2485$.

For the finite comparison, raise the multiplier from one to $e^{.1}$ and keep search cost fixed. Advertising expenditure in the fixed-advertising column is the same baseline vector; firms reprice and consumers search optimally in both columns.

With advertising adjusting, $H$ increases advertising expenditure to $.57004882$ and $M$ cuts it to $.16741657$. All prices rise by about $.000031$, whereas they fall by about $.0005$ with advertising expenditure fixed. These are finite endpoint changes; the derivatives above describe the baseline marginal response.

\paragraph{Uniqueness of the illustrated market.}
For the primitives above, the range test also gives an analytical certificate. The price first-order condition and monotonicity of its cutoff enclose every candidate equilibrium in $.94<d:=v-c-u<1.166$. At the lower endpoint every gain is below $.296112$; at the upper endpoint every gain exceeds $.296262$. The corresponding priced cutoffs lie in $[-.3795,-.27]$ for $H$, $[.6,.73]$ for $M$, and $[-.09,.026]$ for $L$. These enclosures follow by inserting their endpoints into $(p_i-c)(\beta+h_i)=1$; for example, the two limiting $H$ first-order-condition values at $d=1.166$ are below $.997$ and above $1.006$.

Let $\varphi$ be the standard normal density. The Gaussian tail bound $\varphi(2.05)<.049$ gives $S_H>.837$ and $h_H'\le6.125/.837+1.499^2$. Conditional-price monotonicity gives $h_H>1/.788-.01$ and $h_+<1.499$. Define only for this certificate
\[
 R=\frac{(.788)^{-2}}{(.788)^{-2}+6.125/.837+1.499^2}>.1441086.
\]
For $M$ and $L$, the direct losses exceed $.174$ and $.242$, respectively; their bounds for $d_-$ exceed $.265$ and $.44$. Thus
$\min_iS_i(1+\partial_up_i)\ge.837R$ and
$d_-\ge(1/.788-.01)R-.01$. The Gaussian gain formula gives $\Delta_G\le.142656157$, and $\Delta_b\le\beta=.01$. Consequently
\[
 \Delta_G\Delta_J<.443325<.482475
 <4\min_iS_i(1+\partial_up_i).
\]
Strict entry holds since $u>1.834$. The test covers every configuration of positive advertisers and every fixed advertising history. A type multiplier leaves conditional prices and these bounds unchanged, so the certificate applies throughout the illustrated policy interval.

\section{Proof of Proposition~\ref{prop:design-positive-beta}}\label{app:platform-design}

\begin{proof}
We express receipts and implementation in terms of transaction shares, then construct an improvement at constant consumer surplus. Throughout this proof, write $c_\tau=c/(1-\tau)$ for effective marginal cost and $\Gamma=\gamma(1-\tau)$ for the advertising share of transaction returns.

\paragraph{Receipts and advertising participation.}

Let $x_i=u+p_i-v_i$ be the acceptance cutoff, $m_i=p_i-c_\tau$ the markup above effective cost, $S(x)=1-F(x)$ the acceptance probability, and $G(x)=\int_x^\infty S(t)\,\dd t$ the gain from another visit. Write $r$ for the high-quality visit share and $\Delta=v_H-v_L>0$ for the quality gap. Conditional pricing and advertising give
\begin{equation}
 m_i=\frac1{\beta+h(x_i)},\qquad
 \omega_H=\frac{rS(x_H)}{rS(x_H)+(1-r)S(x_L)},\qquad
 z_i=\left[\frac{\Gamma\omega_i m_i}{\alpha_i}-\kappa\right]_+,
 \label{eq:design-adcorners}
\end{equation}
where $\omega_L=1-\omega_H$ and $[x]_+=\max\{x,0\}$. The advertising formula includes the first-order inequality at zero. Cash receipts at every participation pattern are
\begin{equation}
 R=\tau c_\tau+\tau\sum_i\omega_i m_i
 +\sum_i[\Gamma\omega_i m_i-\alpha_i\kappa]_+.
 \label{eq:design-revenue-corners}
\end{equation}
When both types advertise, this is~\eqref{eq:platform-revenue-interior}. Relative scores recover the quality weight implementing an outcome:
\begin{equation}
 k\Delta=\log\frac{r(1-\alpha_H)}{(1-r)\alpha_H}
 +\beta(p_H-p_L)-\gamma\log\frac{1+z_H/\kappa}{1+z_L/\kappa}.
 \label{eq:design-implementation}
\end{equation}
Conditional pricing, advertising, search, and~\eqref{eq:design-implementation} with $k\geq0$ characterize implementable active outcomes.

The conditional pricing equation implies $x_H<x_L$ and $m_H\geq m_L$: the function $x-1/[\beta+h(x)]$ is strictly increasing. High-quality sellers also advertise at least as much as low-quality sellers.\label{lem:design-visit-bound}
To establish this ordering, write $m(x)=1/[\beta+h(x)]$ for the conditional markup and $S(x)=1-F(x)$. Apart from common factors, a seller's coefficient on its advertising multiplier is $e^{kv_i}e^{-\beta m(x_i)}m(x_i)S(x_i)$, where $x_i=u+p_i-v_i$. Its logarithmic derivative with respect to the cutoff is
\[
 \frac{\dd}{\dd x}\log[e^{-\beta m(x)}m(x)S(x)]
 =h(x)[m'(x)-1]<0.
\]
Since $x_H<x_L$ and $k\geq0$, the high-quality coefficient is larger. Strict concavity of the advertising problem gives $z_H\geq z_L$.

\paragraph{An improvement at constant consumer surplus.}

Substituting $p_i=c_\tau+m_i$ gives $x_i=u+c_\tau+m_i-v_i$. At fixed $u$, differentiation with respect to $\beta$ yields
\begin{equation}\label{eq:design-fixed-utility-price}
 \left.\frac{\partial m_i}{\partial\beta}\right|_u
 =\left.\frac{\partial x_i}{\partial\beta}\right|_u
 =-\frac{m_i^2}{1+h'(x_i)m_i^2}<0.
\end{equation}

The next step determines how purchases must change to keep search equally attractive. Let $\omega$ be the high-quality type's transaction share. A consumer's search equation can be written
\begin{equation}\label{eq:design-transaction-search}
 \omega Q(x_H)+(1-\omega)Q(x_L)=0,
 \qquad Q(x)=\frac{G(x)-s}{S(x)}.
\end{equation}
Indeed, transaction shares and visit shares satisfy
$\omega_i=r_iS(x_i)/\sum_jr_jS(x_j)$, so dividing the original search equation by the conversion probability gives \eqref{eq:design-transaction-search}. Increasing hazard implies $h(x)G(x)/S(x)\leq1$ and hence
\[
 Q'(x)=-1+h(x)\frac{G(x)-s}{S(x)}<0.
\]
Thus $Q(x_H)>Q(x_L)$. Holding $u$ constant, the share required by \eqref{eq:design-transaction-search} satisfies
\begin{equation}\label{eq:design-fixed-utility-composition}
 \left.\frac{\partial\omega}{\partial\beta}\right|_u
 =-\frac{\omega Q'(x_H)x_{H,\beta}
 +(1-\omega)Q'(x_L)x_{L,\beta}}
 {Q(x_H)-Q(x_L)}<0.
\end{equation}
Equations \eqref{eq:design-fixed-utility-price}--\eqref{eq:design-fixed-utility-composition} define a continuous nearby outcome path: reducing $\beta$ raises both markups and the high-quality transaction share while keeping $u$ unchanged. For sufficiently small changes the share remains strictly between zero and one. Its average markup
$\overline m=\omega m_H+(1-\omega)m_L$ strictly rises, since
\[
 d\overline m=(m_H-m_L)\,d\omega
              +\omega\,dm_H+(1-\omega)\,dm_L>0.
\]
The strict rise follows from the within-type markup increases even if a weakly increasing hazard makes $m_H=m_L$.

The participation ordering established above leaves three cases for advertising receipts:

\[
 \Ads=[\Gamma\omega m_H-\alpha_H\kappa]_+
     +[\Gamma(1-\omega)m_L-(1-\alpha_H)\kappa]_+.
\]
When both types advertise, this equals $\Gamma\overline m-\kappa$ and rises along the proposed path. When only the high-quality type advertises, it equals $\Gamma\omega m_H-\alpha_H\kappa$ and also rises. When neither advertises, receipts are zero; any entry begins with the high-quality type. These expressions agree at participation boundaries, so advertising receipts cannot fall along a sufficiently short path, including a path through a boundary. Commission receipts $\tau(c_\tau+\overline m)$ strictly rise.

\paragraph{Implementation and consumer participation.}
To implement the improvement, recover visits as $r=(\omega/S(x_H))/[\omega/S(x_H)+(1-\omega)/S(x_L)]$ and the quality weight from~\eqref{eq:design-implementation}. All terms are continuous, including when a seller enters or leaves advertising. Starting from $k>0$, a sufficiently short adjustment remains feasible. The resulting rule implements the constructed equilibrium, preserves consumer surplus $\CS=u+s$, and increases cash receipts. This proves dominance among implementable outcomes.

For rules evaluated through a locally unique continuation, the constructed outcome lies on that continuation, proving the rule-choice claim. The implicit-function theorem supplies local uniqueness at a regular equilibrium on a fixed advertising branch. At advertising boundaries, the outcome construction remains continuous and the rule comparison uses the locally unique continuation stipulated in the proposition. The argument starts at finite positive weights.

For endogenous consumer participation, let $N(u+s)>0$ denote the mass of consumers whose outside option is no greater than their common expected search surplus. Consumers differ only in those outside options. Conditional prices are unchanged, and advertising and total cash receipts are
\begin{align*}
 z_i&=\left[\frac{\Gamma N(u+s)\omega_i m_i}{\alpha_i}-\kappa\right]_+,\\
 R&=\tau N(u+s)(c_\tau+\overline m)
 +\sum_i[\Gamma N(u+s)\omega_i m_i-\alpha_i\kappa]_+.
\end{align*}
The adjustment holds $u$ fixed and hence retains the same entrants. When both types advertise, advertising receipts are $\Gamma N(u+s)\overline m-\kappa$; when only high-quality sellers advertise, they are $\Gamma N(u+s)\omega_Hm_H-\alpha_H\kappa$. Both rise along the adjustment, and commission receipts strictly increase. Each entrant's surplus and aggregate consumer welfare are unchanged.

For fixed seller masses, a seller's operating return before advertising is $t_i=(1-\tau)N(u+s)\omega_i m_i/\alpha_i$, and its optimized profit is $t_i-[\gamma t_i-\kappa]_+$. This profit increases with $t_i$. The adjustment raises $t_H$, while the effect on $t_L$ is ambiguous: fewer sales accompany a larger markup. Continuity preserves every strictly slack participation constraint under a small adjustment. This proves the stated extension to slack seller participation constraints.
\end{proof}

\section{Derivation of the joint-score price response}\label{app:nonseparable}

We derive~\eqref{eq:nonseparable-price} and the associated curvature of optimized operating returns. Use the joint score in Section~\ref{sec:extensions}, write $S_i(x)=1-F_i(x)$, and define
\[
 V_i(z,u)=\max_{p\geq c_i}(p-c_i)w_i(p,z)S_i(u+p-v_i).
\]
The numerator $V_i$ includes the full joint score. Let
\[
 A_i=(p_i-c_i)^{-2}+h_i'(u+p_i-v_i)-\partial_{pp}\log w_i(p_i,z)>0
\]
be the negative derivative of the price first-order condition. Differentiating that condition gives $\partial_z p_i=(\partial_{pz}\log w_i)/A_i$, proving~\eqref{eq:nonseparable-price}.

For the curvature calculation, define $B_i(p,z)=(p-c_i)w_i(p,z)S_i(u+p-v_i)$. At its price optimum,
\[
 B_{ipp}=-V_i A_i,\qquad B_{ipz}=V_i\partial_{pz}\log w_i,\qquad
 B_{izz}=V_i\frac{w_{izz}}{w_i}.
\]
The notation $\partial_{pz}$ means the mixed partial derivative in price and advertising. All derivatives hold $u$ fixed and are evaluated at $(p_i(z,u),z)$. The envelope theorem and the derivative of the optimal price give
\begin{equation}
 \frac{V_{izz}}{V_i}
 =\frac{w_{izz}}{w_i}+\frac{(\partial_{pz}\log w_i)^2}{A_i}.
 \label{eq:nonseparable-main}
\end{equation}
The additional curvature term is nonnegative and is zero under separability, proving the comparison in Section~\ref{sec:extensions}.

\section{Proof of the finite-assortment demand approximation}
\label{app:finite}\label{app:finite-foundation}

We prove~\eqref{eq:finite-main} with an explicit uniform error bound. The argument first bounds search duration, then compares optimal finite search with the stationary stopping rule seller by seller.

There are $N$ sellers and a mass $N$ of consumers. For public prices and scores write $X_j=v_j-p_j+\varepsilon_j$, $S_j^X(y)=\Pr(X_j\ge y)$, and $G_j^X(y)=\E(X_j-y)_+$. Matches are independent. Before each paid inspection, consumers decide whether to continue; the next seller is drawn from the remaining set in proportion to $w_j$. Recall is allowed and uninspected products cannot be bought.

Let $f_j^X$ be the density of $X_j$. Assume uniform constants $\underline w,\bar w,L,\delta,H>0$ such that
\begin{equation}
 \begin{gathered}
 \underline w\le w_j\le\bar w,\qquad \sup_y f_j^X(y)\le L,\qquad
 G_N(0)\ge s+\delta,\qquad \sup_jG_j^X(H)\le s/2,\\
 G_N(y)=\frac{\sum_jw_jG_j^X(y)}{\sum_jw_j}.
 \end{gathered}
 \label{eq:finite-bounds}
\end{equation}
Here $\underline w,\bar w$ bound individual scores, $L$ bounds utility densities, $\delta$ is a strict initial search-gain margin, and $H$ is a utility level above which all expected gains are at most $s/2$. Let $\widehat u_N$ be the stationary reservation utility in~\eqref{eq:finite-main}, satisfying $G_N(\widehat u_N)=s$. Define the weighted acceptance probability $\varphi_N(y)=\sum_jw_jS_j^X(y)/\sum_jw_j$, and the aggregate acceptable-score mass $\mathcal T_N=\sum_jw_jS_j^X(\widehat u_N)$. We establish the sharper bound
\begin{equation}
 Nd_j^{\rm opt}=\frac{w_jS_j^X(\widehat u_N)}{\mathcal T_N/N}
 +O\!\left(\frac{\log^3(N+1)}N\right),
 \label{eq:finite-en-demand-limit}
\end{equation}
uniformly over profiles satisfying~\eqref{eq:finite-bounds}, including admissible unilateral price deviations.

\begin{proof}
The entry and tail restrictions give
\[
 \delta\le \widehat u_N<H,\qquad \varphi_N(\widehat u_N)\ge s/(2H),\qquad \mathcal T_N\ge cN.
\]
Here the constant in the last lower bound is positive and depends only on the uniform assumptions and search cost. The entry and tail conditions place the unique positive root between zero and $H$: $G_N(0)>s$ and $G_N(H)\le s/2<s$. Because $-G_N'=\varphi_N\le1$,
\[
 \delta\le G_N(0)-s=\int_0^{\widehat u_N}\varphi_N(y)\,\dd y\le \widehat u_N.
\]
The acceptance bound requires the tail restriction as well. Averaging the pathwise inequality $(X_j-\widehat u_N)_+\le H\mathbf1_{\{X_j\ge \widehat u_N\}}+(X_j-H)_+$ gives
\[
 s\le H\varphi_N(\widehat u_N)+s/2,
 \qquad \mathcal T_N=\Bigl(\sum_jw_j\Bigr)\varphi_N(\widehat u_N)
       \ge\underline w N s/(2H).
\]
Thus the cutoff stays away from the exit option, and aggregate acceptable-score mass remains proportional to market size. Both margins are uniform over the admitted profiles.

\paragraph{Search duration after any history.}
Throughout this proof, $K,c>0$ denote finite constants that may change from one bound to the next; they depend only on the uniform assumptions and $s$, not on $N$, seller identity, or the admitted profile. For a remaining set $R$, let $W_R=\sum_{j\in R}w_j$ be its total score. Let $V_R(x)$ be the optimal expected utility net of future search costs when the best available utility, including exit, is $x\ge0$. Backward induction gives
\[
 V_R(x)=\max\left\{x,-s+\sum_{j\in R}\frac{w_j}{W_R}
       \E V_{R\setminus j}(\max\{x,X_j\})\right\},\quad V_\varnothing(x)=x.
\]
Stop at indifference. At any state, let $T$ be the optimal number of further inspections and $Y_T$ the highest utility available upon stopping, including the initial best $x$. Then
\[
 s\E T\le\E(Y_T-x)\le H+\frac s2\E T,
\]
because pathwise $Y_T-x\le(H-x)_++\sum_{j\text{ inspected}}(X_j-H)_+$. Independence and the tail bound justify conditional summation. Hence $\E T\le2H/s$ after every history, not only initially. If an integer block length $b$ is at least $4H/s$, conditional Markov's inequality bounds the probability of another $b$ inspections by $1/2$. Iterating this conditional bound gives
\[
 \Pr(T>kb)\le2^{-k},\qquad k=0,1,\ldots,
\]
so all duration moments are uniformly bounded. The initial strict entry margin ensures that one paid draw followed by optimal stopping is preferable to immediate exit. Making that first draw free changes its cost, but leaves the subsequent optimal search path unchanged.

\paragraph{Value and stopping decisions near the stationary cutoff.}
We next localize stopping at histories omitting $k\le c_0N$ firms. The removal estimate used in the argument holds on the larger domain with up to $2c_0N$ omissions. Write $G_R(y)=\sum_{j\in R}w_jG_j^X(y)/W_R$. If $R$ omits at most $k\le2c_0N$ sellers, then $W_R\ge\underline w(N-k)$ and the removed numerator is at most $k\bar w(H+s/2)$ for $y\ge0$. Comparing the two numerator--denominator ratios gives
\begin{equation}
 \sup_{y\ge0}|G_R(y)-G_N(y)|\le Kk/N.
 \label{eq:finite-removal}
\end{equation}
The same calculation for survival probabilities bounds their change by $Kk/N$. Since $\varphi_N(\widehat u_N)\ge s/(2H)$, choose a removal fraction $c_0\in(0,1/4)$ sufficiently small that the remaining acceptance probability at $\widehat u_N$ is at least $a:=s/(4H)>0$ after at most $2c_0N$ removals. Put $v(x)=\max\{x,\widehat u_N\}$, the stationary comparison value. Its conditional expected increment from a draw is $G_R(\max\{x,\widehat u_N\})$. Starting with $k\le c_0N$ omissions, the first $c_0N$ further draws leave at most $2c_0N$ sellers removed, so the preceding removal and acceptance bounds apply throughout that phase. Its net increment after $t$ draws is at most $K(k+t)/N$. Telescoping along the optimal path gives
\[
 V_R(x)-v(x)\le\frac KN\E\sum_{t<T}(k+t)
                  +K\E[T\mathbf1_{\{T>c_0N\}}]\le K(k+1)/N.
\]
In the late part of this bound, the absolute expected one-draw drift is uniformly bounded: $G_R(y)\le H+s/2$ for $y\ge0$. The exponential duration tail therefore controls the residual after $c_0N$ draws, whereas bounded first and second moments control the displayed early sum.

For a lower bound at $x<\widehat u_N$, inspect until an offer reaches $\widehat u_N$ or the market is exhausted. For its count $T^{\rm thr}$, the acceptance bound gives
\[
 \Pr(T^{\rm thr}>t)\le(1-a)^t\quad(t\le c_0N),\qquad
 \Pr(T^{\rm thr}>c_0N)\le e^{-cN}.
\]
The remaining horizon is at most $N$, so this policy too has uniformly bounded moments. Before success its net drift is $G_{R_t}(\widehat u_N)-s$, bounded in absolute value by $K(k+t)/N$ during the early phase. At unsuccessful exhaustion its best offer differs from $v$ by at most $\widehat u_N\le H$, on an exponentially unlikely event. Telescoping and the same early--late decomposition give $V_R(x)\ge v(x)-K(k+1)/N$. For $x\ge \widehat u_N$, immediate stopping proves the lower bound directly.

Let $\mathcal C_R$ denote the continuation expression inside the Bellman maximum. For $k+1\le c_0N$, apply the established value bound at each next-period state and then use~\eqref{eq:finite-removal}. This gives the uniform comparison
\[
 |\mathcal C_R(x)-\mathcal C_N^0(x)|\le e_k:=K(k+1)/N,
 \qquad
 \mathcal C_N^0(x)=
 \begin{cases}
 \widehat u_N,&x\le \widehat u_N,\\
 x+G_N(x)-s,&x\ge \widehat u_N.
 \end{cases}
\]
Below $\widehat u_N$, the benchmark continuation advantage is $\widehat u_N-x$, so the optimal policy continues whenever $x<\widehat u_N-2e_k$. To obtain a uniform stopping margin above $\widehat u_N$, put $a_0=s/(2H)$. The density bound implies $|\varphi_N(y)-\varphi_N(\widehat u_N)|\le L|y-\widehat u_N|$. Hence, for $0\le d\le a_0/(2L)$,
\[
 s-G_N(\widehat u_N+d)=\int_{\widehat u_N}^{\widehat u_N+d}\varphi_N(y)\,\dd y
 \ge a_0d/2.
\]
For sufficiently small $(k+1)/N$, the optimal continuation advantage is therefore strictly negative at $x=\widehat u_N+4e_k/a_0$. Backward induction also makes $V_R$ increasing and 1-Lipschitz. Thus, for $y>x$,
\[
 [\mathcal C_R(y)-y]-[\mathcal C_R(x)-x]
 \le-\sum_{j\in R}\frac{w_j}{W_R}
               \E\min\{(X_j-x)_+,y-x\}\le0.
\]
A negative continuation advantage just above the cutoff therefore stays negative at every larger current best. Enlarging $K$ to cover both margins, optimal and fixed-cutoff decisions agree outside
\[
 [\widehat u_N-K(k+1)/N,\ \widehat u_N+K(k+1)/N].
\]

\paragraph{Individual demand under the fixed-cutoff policy.}
Let $d_j^{\rm thr}$ denote seller $j$'s purchase probability under the fixed-cutoff policy. For that policy, independent exponential clocks of rates $w_j$ generate the weighted permutation. Conditional on $X_j\ge \widehat u_N$, seller $j$ is the first acceptable offer with probability $w_j\E(1/Z_j)$, where
\[
 Z_j=w_j+\sum_{k\ne j}w_k\mathbf1_{\{X_k\ge \widehat u_N\}},\qquad
 \E Z_j=\mathcal T_N+O(1),\quad\operatorname{Var}(Z_j)=O(N).
\]
Bounded-summand concentration gives $\Pr(Z_j<\E Z_j/2)\le e^{-cN}$. Using $Z_j\ge\underline w$ and the exact expansion
\[
 Z_j^{-1}=m^{-1}-(Z_j-m)m^{-2}+(Z_j-m)^2/(m^2Z_j),\quad m=\E Z_j,
\]
yields $\E(1/Z_j)=1/\mathcal T_N+O(N^{-2})$. Indeed, on $Z_j\ge m/2$ the expected remainder is at most $2\operatorname{Var}(Z_j)/m^3=O(N^{-2})$; on its complement the lower bound on $Z_j$ and concentration make its expectation exponentially small. Exhaustion has probability at most $\prod_k[1-S_k^X(\widehat u_N)]\le e^{-cN}$. Thus $d_j^{\rm thr}=w_jS_j^X(\widehat u_N)/\mathcal T_N+O(N^{-2})$.

\paragraph{The difference in individual purchase probabilities.}
Couple both policies on the same permutation and matches, and truncate after $L_N=A\log(N+1)$ draws, where $A>0$ is a fixed truncation constant chosen sufficiently large. Their duration bounds make the truncation probability $O(N^{-3})$ for large fixed $A$. For large $N$, these histories satisfy $L_N+1\le c_0N$, so the localization bound applies. Earlier disagreement requires some observed utility within $K(L_N+1)/N$ of $\widehat u_N$; this interval excludes zero. Changing seller $j$'s purchase also requires its inclusion among these labels, an event of probability $O(L_N/N)$. Here the prefix is the first $L_N$ labels of the complete weighted permutation, whether or not either policy reaches them. This permutation is independent of all matches. Conditional on any such prefix, the utility draws still have their original independent distributions. The density bound and a union bound make the near-cutoff event's conditional probability $O(L_N^2/N)$, uniformly in the prefix. Averaging this bound only over prefixes containing $j$ multiplies it by their inclusion probability $O(L_N/N)$; no independence between the two events is assumed. Therefore
\[
 |d_j^{\rm opt}-d_j^{\rm thr}|=O(L_N^3/N^2)+O(N^{-3}).
\]
Combining the two bounds proves~\eqref{eq:finite-en-demand-limit}, and hence~\eqref{eq:finite-main}.
\end{proof}

\label{end:appendix}
\end{document}